\documentclass[reprint,aps,prd,amsmath,amssymb,nofootinbib,superscriptaddress,floatfix]{revtex4-2}
\usepackage{graphicx}
\usepackage{dcolumn}
\usepackage[utf8]{inputenc}
\usepackage{xcolor}
\usepackage[unicode=true,colorlinks,linkcolor=blue,anchorcolor=blue,urlcolor=blue,citecolor=blue,breaklinks=true]{hyperref}
\usepackage{epstopdf}
\usepackage{orcidlink}
\newcommand{\ii}{\mathrm{i}\,}

\newcommand{\pararrow}{\mathord{\buildrel{\lower3pt\hbox{$\scriptscriptstyle\leftrightarrow$}}\over {\partial}}} 
\newcommand{\pararrowk}[1]{\mathord{\buildrel{\lower3pt\hbox{$\scriptscriptstyle\leftrightarrow$}}\over {\partial}\hspace*{-0.18em}{}^#1}\hspace*{-0.18em} \,} 

\usepackage{bm}
\usepackage{slashed}

\newcommand{\qfnu}{\affiliation{College of Physics and Engineering, Qufu Normal University, Qufu 273165, China}}

\newcommand{\imp}{\affiliation{State Key Laboratory of Heavy Ion Science and Technology, Institute of Modern Physics, Chinese Academy of Sciences, Lanzhou 730000, China}}
\newcommand{\snst}{\affiliation{School of Nuclear Science and Technology, University of Chinese Academy of Sciences, Beijing 101408, China}}
\newcommand{\scnt}{\affiliation{Southern Center for Nuclear-Science Theory (SCNT), Institute of Modern Physics, Chinese Academy of Sciences, Huizhou 516000, Guangdong, China}}
\newcommand{\hnnu}{\affiliation{Institute of Particle and Nuclear Physics, Henan Normal University, Xinxiang 453007, China}}

\begin{document}
	
	\title{Study of $\chi_{cJ}\to \eta \eta \eta^\prime$ via intermediate charmed meson loop mechanisms and  its implications for non-observation of $\eta_1(1855)$ in $\chi_{cJ}$ decays}
	
	\author{Xin-Ru Wang} \qfnu
	\author{Shu-Qi Wang} \qfnu
	\author{Shi-Dong Liu\,\orcidlink{0000-0001-9404-5418}} \email{liusd@qfnu.edu.cn}\qfnu
	\author{Qi Wu\,\orcidlink{0000-0002-5979-8569}}\email{wuqi@htu.edu.cn} \hnnu
    \author{Gang Li\,\orcidlink{0000-0002-5227-8296}} \email{gli@qfnu.edu.cn} \qfnu	
    \author{Ju-Jun Xie\,\orcidlink{0000-0001-9888-5924}} \email{xiejujun@impcas.ac.cn} \imp \snst \scnt
	
	\begin{abstract}
    Recently, the BESIII Collaboration reported the first observation of the decays $\chi_{cJ} \to \eta \eta \eta^\prime$ in order to search for the $1^{-+}$ exotic state $\eta_1(1855)$. A partial wave analysis of the $\eta \eta^\prime$ invariant mass spectrum shows no significant signal for the $\eta_1(1855)$. In this work, we, using an effective Lagrangian approach, investigate the processes $\chi_{cJ} \to \eta \eta \eta^\prime$ via the box and triangle loops involving charmed mesons and the scalar meson $f_0(1500)$. Our calculations reproduce well the experimental branching fractions of $\chi_{cJ} \to \eta \eta \eta^\prime$. Furthermore, we present the predictions of the relevant invariant mass spectra of $\eta \eta^\prime$ and $\eta \eta$ produced in the $\chi_{c1}$ decay, which seem overall consistent with the BESIII measurements. In the present model, the decay $\chi_{c1} \to \eta \eta \eta^\prime$ is dominated by the triangle and box loop contributions. The consistency between our theoretical results and the BESIII measurements sheds light on the underlying decay mechanism of the $\chi_{cJ}$ decaying into light mesons and might be helpful to understand the absence of the $\eta_1(1855)$ signal in the decay channels $\chi_{cJ} \to \eta \eta \eta^\prime$.		
 \end{abstract}
	
	\date{\today}
	
	\maketitle
	\section{Introduction}\label{sec:intro}
	In 2022, the BESIII Collaboration established for the first time an isoscalar state with exotic quantum numbers $J^{PC}=1^{-+}$, denoted as $\eta_1(1855)$, in the process $J/\psi \to \gamma \eta_1(1855)\to \gamma \eta \eta^\prime$~\cite{BESIII:2022riz,BESIII:2022iwi,Maon:2024ifr} with a statistical  significance larger than $19\sigma$. 
	The mass and width of the $\eta_1(1855)$ were measured to be $(1855 \pm 9^{+6}_{-1})$ MeV and $(188 \pm 18^{+3}_{-8})$ MeV, respectively. 
	According to the PDG~\cite{ParticleDataGroup:2024cfk}, the $\eta_1(1855)$ apparently ever leave a trace in the central $pp$ collision data at 450 GeV/c by the WA102 Collaboration~\cite{WA102:1999hsn} and the $\pi^- p$ charge-exchange reaction in the SERP experiments~\cite{Alde:1991qz}, but showing a larger mass of $ \gtrsim 1.9$ GeV and a smaller width between $(90 \pm 35)$ and $(141 \pm 41)$ MeV. 

	The $\eta_1(1855)$ could be considered as the isoscalar partner of the isovector states $\pi_1(1400)$, $\pi_1(1600)$, and $\pi_1(2015)$ \cite{Qiu:2022ktc,Chen:2022qpd}, which are the only three exotic-quantum-number candidates observed in experiments to date \cite{ParticleDataGroup:2024cfk,Meyer:2015eta}. 
	Hence, the discovery of the $\eta_1(1855)$ enhances the existence of the $1^{-+}$ exotic state family. 
	The $\pi_1(1400)$ and $\pi_1(1600)$ states have been widely studied both experimentally and theoretically over the past few decades \cite{Meyer:2015eta} (and references therein and thereto). 
	Experimentally, the $\pi_1(1400)$ was mainly observed in the $\eta \pi$ decay channel and has a much lower mass than the lattice predictions for light-quark hybrid mesons ranging from 1.6 to 2.2 GeV~\cite{Meyer:2015eta,Dudek:2013yja,Klempt:2007cp}, while the $\pi_1(1600)$ was observed in various decay channels, such as $\eta' \pi$ \cite{VES:1993scg,Khokhlov:2000tk,E852:2004gpn,CLEO:2011upl,COMPASS:2014vkj,Zaitsev:2000rc}, $b_1\pi$ \cite{Zaitsev:2000rc,E852:2004rfa,Amelin:2005ry,Baker:2003jh}, $f_1(1285) \pi$ \cite{E852:2004gpn,Amelin:2005ry}, and $\rho \pi$ \cite{E852:1998mbq,Zaitsev:2000rc,Chung:2002pu,COMPASS:2009xrl,COMPASS:2018uzl,COMPASS:2021ogp}, with a mass in good agreement with the naive extrapolation of the lattice results to the physical pion mass \cite{Meyer:2015eta}. 
	The higher mass state $\pi_1(2015)$ was reported only by the E852 experiments \cite{E852:2004rfa,E852:2004gpn}, but not confirmed by other experiments. 
	The coupled channel analyses \cite{JPAC:2018zyd,Kopf:2020yoa} indicates the presence of only one $\pi_1$ meson near 1.6 GeV, namely $\pi_1(1600)$, standing as a better $1^{-+}$ isovector candidate \cite{ParticleDataGroup:2024cfk}. The $\pi_1(2015)$ could be  considered as a radially excited state of the $\pi_1(1600)$ \cite{Dudek:2010wm,Esmer:2025xss}.
	The theoretical investigations on the nature of the $\pi_1$ states have been carried out in various models (see Refs.~\cite{Meyer:2015eta,Klempt:2007cp} for a review of hybrid mesons). 
	%

	The $\eta_1(1855)$ is widely interpreted as hybrid mesons (quark-antiquark-gluon) \cite{Qiu:2022ktc,Chen:2022qpd,Shastry:2022mhk,Chen:2022isv,Chen:2023ukh,Shastry:2023ths,Liang:2024lon,Zhang:2025xee}, with the $s\bar{s}g$ hybrid assignment being a notable example \cite{Chen:2022qpd,Shastry:2022mhk,Chen:2023ukh}. 
	Apart from the hybrid meson interpretation, the $\eta_1(1855)$ has also been proposed as a hadronic molecule composed of $K \bar{K}_1(1400)+ \mathrm{c.c}$~\cite{Dong:2022cuw,Yang:2022rck,Wang:2022sib,Yu:2022lwl} in an $S$ wave, in view of its mass just $\sim 40$ MeV below the $K \bar{K}_1(1400)$ threshold. 
	A tetraquark picture with the $[1_c]_{s\bar{s}}\otimes[1_c]_{q\bar{q}}$ configuration has also been studied using QCD sum rules \cite{Wan:2022xkx,Qiao:2023dhf}. 
	The $\eta_1(1855)$ could be interpreted as a dynamically generated resonance through the pseudoscalar-axial vector meson interactions within the chiral unitary approach \cite{Yan:2023vbh}.
	A mixed state of hybrid meson and tetraquark state might also be possible \cite{Tan:2025ahx}.
	The decay and production properties of the $\eta_1(1855)$ have been extensively studied using different models in the literature cited above.

	\begin{figure*}[htbp]
		\centering
		\includegraphics[width=0.96\linewidth]{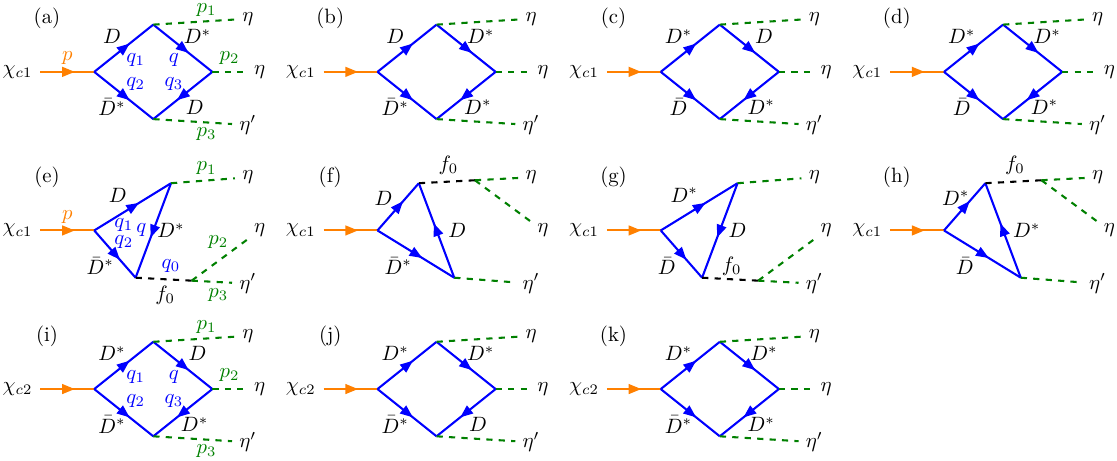}
		\caption{Feynman diagrams for the processes $\chi_{c1} \to \eta \eta \eta^\prime$ [(a)--(h)] and $\chi_{c2} \to \eta \eta \eta^\prime$ [(i)--(k)]. As indicated, both box loops (a)--(d) and triangle loops (e)--(h) with the intermediate $f_0(1500)$ contribute to the $\chi_{c1}$ decay, while for $\chi_{c2}$ decay, only the box loops (i)--(k) contribute. The relevant kinematics [$p, p_{1(2,3)}, q, q_{0(1,2,3)}$] are explicitly indicated in the graphs. The loops due to the charge conjugation of the charmed mesons and the exchange symmetry of the final pseudoscalar mesons are not shown, but are included in the calculations.}
		\label{fig:feyndiagschic2etas}
	\end{figure*}

	Motivated by the $\eta_1(1855)$ observation, some theoretical and experimental efforts have been paid to search for the chamonium-like or fully-charmed $1^{-+}$ state (in the heavy flavor sector) have been performed \cite{Zhang:2025gmm,Shi:2023sdy,Xu:2025epx,BESIII:2025vea,Wan:2025sae,BESIII:2025bez}. 
	However, the $\eta_1(1855)$ internal structure is still not fully understood, and any of the above interpretations seems possible \cite{Huang:2022tpq,Wang:2022sib}, needing further experimental and theoretical efforts. 
	Since the $\eta_1(1855)$ was first observed in the invariant mass spectrum of $\eta \eta^\prime$ of the radiative decay $J/\psi \to \gamma \eta \eta^\prime$ \cite{BESIII:2022riz,BESIII:2022iwi,Maon:2024ifr}, it is expected to observed it in other processes in which the $\eta\eta^\prime$ could be created. 
	Hence, the BESIII Collaboration recently performed a search for the $\eta_1(1855)$ in the decay processes $\chi_{cJ}(1P) \to \eta \eta \eta^\prime$ ($J=0,1,2$) \cite{BESIII:2025izw}.
	Nevertheless, no significant signal for the $\eta_1(1855)$ was found in the $\eta \eta^\prime$ invariant mass spectrum. 
	In terms of a partial wave analysis, the structure in the $\eta\eta^\prime$ mass spectrum is mainly attributed to the $f_0(1500)$ and the $\eta\eta$ mass spectrum is primarily the $0^{++}$ phase space. 
	The measured upper limit of the product branching fraction $\mathcal{B}[\chi_{c1} \to \eta_1(1855) \eta]\cdot \mathcal{B}[\eta_1(1855) \to \eta \eta^\prime]$ at the 90\% confidence level were determined to be $9.79 \times 10^{-5}$.
	Non-observation of the $\eta_1(1855)$ in this channel may be due to the limited statistics, and more data are needed to draw a definite conclusion. Perhaps, this channel is not favorable for producing the $\eta_1(1855)$.
	
	In this work, we restrict our attention to the decay processes $\chi_{cJ}(1P) \to \eta \eta \eta^\prime$ ($J=1,2$) to understand the underlying mechanism of these processes and the non-observation of the $\eta_1(1855)$ signal in this channel. 
	In the calculations, we consider the box and triangle loop mechanisms involving charmed mesons and the scalar $f_0(1500)$. 
	%
    Our calculations show that the triangle loop contributes somewhat more than the box loop to the decay $\chi_{c1}\to \eta \eta \eta^\prime$ with present model parameters.
	The overall consistency of our theoretical results with the BESIII measurements sheds light on the underlying decay mechanism of the $\chi_{cJ}$ decaying into light mesons and helps explain the absence of the $\eta_1(1855)$ signal in the decay channel $\chi_{cJ} \to \eta \eta \eta^\prime$.
	
	In the following, we first give in Sec. \ref{sec:lags} the effective Lagrangians. Then, in Sec. \ref{sec:results} the numerical results and discussion are described in detail. Finally, a summary is given in Sec. \ref{sec:summary}.

\section{Lagrangian Formalism}\label{sec:lags}

	In the present study, we assume the processes $\chi_{cJ} \to \eta \eta \eta^\prime$ to occur via the charmed meson loops. However, under the leading order Lagrangians that we adopt in the following, there is no loop contribution to the $\chi_{c0} \to \eta \eta \eta^\prime$ decay, and hence we will not consider this process. For the $\chi_{c1} \to \eta \eta \eta^\prime$ decay, both box and triangle loops with the intermediate $f_0(1500)$ contribute, while for the $\chi_{c2} \to \eta \eta \eta^\prime$ decay, only box loops contribute. The consideration of the triangle loops with the intermediate $f_0(1500)$ for the $\chi_{c1}$ decay is motivated by the BESIII partial wave analysis of the $\eta \eta^\prime$ invariant mass spectrum, which indicates that the structure in the $\eta \eta^\prime$ mass spectrum is mainly attributed to the $f_0(1500)$~\cite{BESIII:2025izw}. The Feynman diagrams for the processes $\chi_{c1} \to \eta \eta \eta^\prime$ and $\chi_{c2} \to \eta \eta \eta^\prime$ are shown in Fig. \ref{fig:feyndiagschic2etas}.
	
	To calculate the decay amplitudes of the processes $\chi_{cJ} \to \eta \eta \eta^\prime$, we need to introduce the effective Lagrangians relevant to the interactions among the involved mesons. Under the heavy quark limit, the effective Lagrangians for the interactions of $\chi_{cJ}$ with charmed mesons can be written as \cite{Colangelo:2003sa,Casalbuoni:1996pg,Li:2021jjt,Liu:2025sjz}
	\begin{align}
		\mathcal{L}_{\chi} &= g_{\chi{c0}\mathcal{D}\mathcal{D}}\chi_{c0}\bar{\mathcal{D}}^\dagger\mathcal{D}^\dagger - g_{\chi_{c0}\mathcal{D}^\ast\mathcal{D}^\ast}\bar{\mathcal{D}}^{\ast\dagger\mu}\mathcal{D}^{\ast\dagger}_\mu\nonumber\\
		&+ g_{\chi_{c1}\mathcal{D}\mathcal{D}^\ast} \chi_{c1}^\mu (\bar{\mathcal{D}}^{\ast\dagger}_\mu\mathcal{D}^\dagger - \bar{\mathcal{D}}^\dagger\mathcal{D}^{\ast\dagger}_\mu)\nonumber\\
		& + g_{\chi_{c2}\mathcal{D}^\ast\mathcal{D}^\ast}  \chi_{c2}^{\mu\nu}(\bar{\mathcal{D}}^{\ast\dagger}_\mu\mathcal{D}^{\ast\dagger}_\nu+\bar{\mathcal{D}}^{\ast\dagger}_\nu\mathcal{D}^{\ast\dagger}_\mu)\,.
	\end{align}
	Here $\mathcal{D}^{(\ast)}$ and $\bar{\mathcal{D}}^{(\ast)}$ represent the charmed meson triplets, which are defined as $\mathcal{D}^{(\ast)}=(D^{(\ast)0},D^{(\ast)+},D_s^{(\ast)+})$ and $\bar{\mathcal{D}}^{(\ast)}=(\bar{D}^{(\ast)0},D^{(\ast)-},D_s^{(\ast)-})$. The coupling constants $g_{\chi_{cJ}\mathcal{D}^{(\ast)}\bar{\mathcal{D}}^{(\ast)}}$ are linked each other by a global coupling constant $g_1$:
	\begin{align}
		g_1&=\frac{g_{\chi_{c0}\mathcal{D}\mathcal{D}}}{2\sqrt{3m_{\chi_{c0}}}m_{\mathcal{D}}} = \frac{\sqrt{3}g_{\chi_{c0}\mathcal{D}^\ast\mathcal{D}^\ast}}{2\sqrt{m_{\chi_{c0}}}m_{\mathcal{D}^\ast}}\nonumber\\
		&=\frac{g_{\chi_{c1}\mathcal{D}\mathcal{D}^\ast}}{2\sqrt{2m_{\chi_{c1}}m_{\mathcal{D}}m_{\mathcal{D}^\ast}}} =\frac{g_{\chi_{c2}\mathcal{D}^\ast\mathcal{D}^\ast}}{2\sqrt{m_{\chi_{c2}}}m_{\mathcal{D}^\ast}}\,.
	\end{align}
	In terms of the vector meson dominance, $g_1= -\sqrt{m_{\chi_{c0}}/6}/f_{\chi_{c0}}$ \cite{Colangelo:2003sa,Deandrea:2003pv,Mehen:2015efa}, where $f_{\chi_{c0}}$ are the $\chi_{c0}$ decay constants with the value of $f_{\chi_{c0}} = (510\pm 10) $ MeV estimated in the framework of the QCD sum rules \cite{Colangelo:2002mj} and $m_{\chi_{c0}} = (3414.71\pm 0.30)$ MeV \cite{ParticleDataGroup:2024cfk}. Then, we can get $g_1 = -(1.48\pm 0.03)$ GeV$^{-1/2}$. However, this model predicted value of $g_1$ seems too large in comparison with the value of $\left|g_1\right|\lesssim 0.53^{+1.3}_{-0.13}~\mathrm{GeV}^{-1/2}$ extracted from the experimental data of $\chi_{c1}(3872) \to \chi_{cJ} \pi^0$ \cite{Jia:2025xil,Wu:2025crk}, where the $\chi_{c1}(3872)$ is considered as the $D\bar{D}^\ast$ molecule. Nevertheless, we shall use the average value of $g_1 = -1.0~\mathrm{GeV}^{-1/2}$ in the following.

	The effective Lagrangian for the interactions of the charmed mesons with the light pseudoscalar mesons is \cite{Colangelo:2003sa,Casalbuoni:1996pg,Li:2021jjt,Liu:2025sjz}
	\begin{align}
		\mathcal{L}_\mathcal{P}&=\ii g_{\mathcal{D}\mathcal{D}^\ast\mathcal{P}}(\mathcal{D}_b \partial_\mu\mathcal{P}_{ba}\mathcal{D}^{\ast\dagger \mu}_{a}-\mathcal{D}^{\ast}_b\partial_\mu\mathcal{P}_{ba}\mathcal{D}^{\dagger\mu}_a)\nonumber\\
		&-\frac{1}{2}g_{\mathcal{D}^\ast\mathcal{D}^\ast\mathcal{P}} \epsilon_{\mu\nu\alpha\beta}\mathcal{D}^{\ast\mu}_b\partial^\nu\mathcal{P}_{ba}\pararrowk{\alpha}\mathcal{D}^{\ast\dagger\beta}_{ba}\,,
	\end{align}
	where $\mathcal{P}$ is a $3\times 3$ matrix of the light pseudoscalar mesons, which is defined as
	\begin{align}
		\mathcal{P}=\begin{pmatrix}
			\frac{\pi^0}{\sqrt{2}}+\alpha \eta +\beta \eta^\prime & \pi^+ & K^+ \\
			\pi^- & -\frac{\pi^0}{\sqrt{2}}+\alpha \eta +\beta \eta^\prime & K^0 \\
			K^- & \bar{K}^0 & \gamma \eta +\delta \eta^\prime
		\end{pmatrix}\,.
	\end{align}
	The physical states $\eta$ and $\eta^\prime$ are the mixtures of the flavor eigenstates \cite{Amsler:1995td}:
	\begin{align}
		\begin{pmatrix}
			\eta \\
			\eta^\prime
		\end{pmatrix}=
		\begin{pmatrix}
			\cos \phi & -\sin \phi \\
			\sin \phi & \cos \phi
		\end{pmatrix}
		\begin{pmatrix}
			|n\bar{n}\rangle \\
			|s\bar{s}\rangle
		\end{pmatrix}\,,
	\end{align}
	with $|n\bar{n}\rangle = (|u\bar{u}\rangle + |d\bar{d}\rangle)/\sqrt{2}$. The $\phi$ is related to the usual octet-singlet mixing angle $\theta_{\mathrm{P}}$ by $\phi = \theta_{\mathrm{P}} + \arctan\sqrt{2}$. By fitting the DM2 measurements of the branching fractions of $J/\psi \to \mathrm{vector}+\mathrm{pseudoscalar}$ \cite{DM2:1988bfq}, the value of $\theta_{\mathrm{P}}$ is estimated to be $(-19.1\pm 1.4)^\circ$. With this value of $\theta_{\mathrm{P}}$, we obtain 
	\begin{align}
		\alpha &= \frac{\cos \phi}{\sqrt{2}} = 0.57\,,\nonumber\\
		\beta &= \frac{\sin \phi}{\sqrt{2}} = 0.41\,,\nonumber\\
		\gamma &= -\sin \phi = -0.58\,,\nonumber\\
		\delta &= \cos \phi = 0.81\,.
	\end{align}
	 The coupling constants $g_{\mathcal{D}\mathcal{D}^\ast\mathcal{P}}$ and $g_{\mathcal{D}^\ast\mathcal{D}^\ast\mathcal{P}}$ have the following relations:
	 \begin{equation}	g_{\mathcal{D}^\ast\mathcal{D}^\ast\mathcal{P}}=\frac{g_{\mathcal{D}\mathcal{D}^\ast\mathcal{P}}}{\sqrt{m_{\mathcal{D}}m_{\mathcal{D}^\ast}}} = \frac{2g}{f_\pi}\,,
	 \end{equation}
	 where $f_\pi = (130.2\pm 1.2)~\mathrm{MeV}$ is the pion decay constant \cite{ParticleDataGroup:2024cfk} and $g$ is a dimensionless coupling constant with the value of $g=0.57\pm 0.01$ extracted from the experimental data of $D^{\ast +} \to D^0 \pi^+(D^+\pi^0)$ \cite{ParticleDataGroup:2024cfk,Liu:2025sjz,Liu:2025bjm}.

	 For the coupling of the scalar $f_0(1500)$ to the charmed mesons, we adopt the following Lagrangian: \cite{Meng:2008dd,Bai:2022cfz,Chen:2011qx,Chen:2015bma}
	\begin{align}
		\mathcal{L}_{f_0} = g_{f_0\mathcal{D}\mathcal{D}}f_0 \mathcal{D}\mathcal{D}^\dagger-g_{f_0 D^\ast D^\ast} f_0 \mathcal{D}^\ast_\mu \mathcal{D}^{\ast\mu\dagger} \,,
	\end{align}
	with the coupling constants being
	\begin{equation}\label{eq:gf0}
		\frac{g_{f_0\mathcal{D}\mathcal{D}}}{m_{\mathcal{D}}} = \frac{g_{f_0\mathcal{D}^\ast\mathcal{D}^\ast}}{m_{\mathcal{D}^\ast}} = g_{f_0}.
	\end{equation}
	The coupling $g_{f_0}$ is not well known and we shall estimate it in the following section. The coupling of the scalar meson $f_0(1500)$ to the pseudoscalar light mesons $\eta$ or $\eta^\prime$ is described by the following Lagrangian:
	 \begin{align}
		\mathcal{L}_{\eta}&= g_{f_0\eta\eta}f_0\eta \eta + g_{f_0\eta\eta^\prime}f_0\eta \eta^\prime\,.
	\end{align}
	The coupling constants $g_{f_0\eta\eta}$ and $g_{f_0\eta\eta^\prime}$ can be determined by the partial decay widths of $f_0(1500) \to \eta \eta$ and $f_0(1500) \to \eta \eta^\prime$, respectively. Using the current world average values of the mass $m_{f_0} = (1522\pm 25)~\mathrm{MeV}$, total width $\Gamma_{f_0} = (108\pm 33)~\mathrm{MeV}$, and branching fractions $\mathcal{B}[f_0(1500) \to \eta \eta] = (6.0\pm 0.9)\%$ and $\mathcal{B}[f_0(1500) \to \eta \eta^\prime] = (2.2\pm 0.8)\%$ \cite{ParticleDataGroup:2024cfk}, we obtain $g_{f_0\eta\eta} = (1.20\pm 0.20)$ GeV and $g_{f_0\eta\eta^\prime} = (1.14\pm 0.50)$ GeV.

	The general form of the box loop amplitude $\mathcal{M}_{\mathrm{box}}$ is
	\begin{align}\label{eq:boxamp}
	\ii \mathcal{M}_{\mathrm{box}} &= \int \frac{d^4 q}{(2\pi)^4} \frac{\mathcal{V}_{\chi}\mathcal{V}_{\eta}\mathcal{V}_{\eta}\mathcal{V}_{\eta^\prime}}{\mathcal{S}_1\mathcal{S}_2\mathcal{S}_3\mathcal{S}}\nonumber\\
 	&\times \mathcal{F}(q_1,M_1)\mathcal{F}(q_2,M_2)\mathcal{F}(q_3,M_3)\mathcal{F}(q,M)\,.
	\end{align}
	The amplitudes of the triangle loops $\mathcal{M}_{\mathrm{tri}}$ have the following general form:
	\begin{align}\label{eq:triamp}
	\ii \mathcal{M}_{\mathrm{tri}} &= \int \frac{d^4 q}{(2\pi)^4} \frac{\mathcal{V}_{\chi}\mathcal{V}_{\eta}\mathcal{V}_{f_0}\mathcal{V}_{\eta\eta^{(\prime)}}}{\mathcal{S}_1\mathcal{S}_2\mathcal{S}}\frac{\ii}{q_0^2-M_{f_0}^2+\ii M_{f_0}\Gamma_{f_0}} \nonumber\\
	&\times \mathcal{F}(q_1,M_1)\mathcal{F}(q_2,M_2)\mathcal{F}(q,M)\,.
	\end{align}
	In Eqs. \eqref{eq:boxamp} and \eqref{eq:triamp}, $\mathcal{V}_{\chi}$, $\mathcal{V}_{\eta}$, $\mathcal{V}_{\eta^\prime}$, and $\mathcal{V}_{f_0}$ represent the vertices of the mesons specified by the subscripts with the charmed mesons, and $\mathcal{V}_{\eta\eta^{(\prime)}}$ represents the vertex of the scalar $f_0(1500)$ coupling to the $\eta \eta$ or $\eta \eta^\prime$ pair. The symbols $\mathcal{S}_1$, $\mathcal{S}_2$, $\mathcal{S}_3$, and $\mathcal{S}$ denote the propagators of the intermediate mesons in the loops, which have the momentum $q_1$, $q_2$, $q_3$, and $q$ and the mass $M_1$, $M_2$, $M_3$, and $M$, respectively (see Fig. \ref{fig:feyndiagschic2etas}). The form factor $\mathcal{F}(q_i,M_i)$ is phenomenologically introduced to regularize the ultraviolet divergence of the loop integrals and to compensate for the off-shell effects of the exchanged mesons \cite{Guo:2010ak,Wang:2010iq,Wang:2012mf}. The form factor is taken as
	\begin{equation}
	\mathcal{F}(q,M) = \frac{M^2 - \Lambda^2}{q^2 - \Lambda^2}\,,
	\end{equation}
	where $\Lambda$ is the cutoff parameter, which is reparameterized as $\Lambda = M + \alpha \Lambda_{\mathrm{QCD}}$ with $\Lambda_{\mathrm{QCD}} = 220$ MeV and $\alpha$ being a dimensionless parameter of order unity \cite{Cheng:2004ru}. In this work, we will vary the value of $\alpha$ in the range of $0.6\leq \alpha \leq 1.2$ to estimate the uncertainties of the results due to the form factor.
	
	The total invariant decay amplitude of the process $\chi_{cJ} \to \eta \eta \eta^\prime$ can be written as
	\begin{equation}\label{eq:amptot}
		\mathcal{M}_{\mathrm{tot}} = \mathcal{M}_{\mathrm{box}} + e^{\ii \theta} \mathcal{M}_{\mathrm{tri}}\,,
	\end{equation}
	where $\theta$ is the relative phase angle between the box and triangle loop amplitudes. Notice that for the process $\chi_{c2} \to \eta \eta \eta^\prime$, only the box loop amplitude contributes. The differential decay width of the process $\chi_{cJ} \to \eta \eta \eta^\prime$ is given by
	\begin{align}
		d\Gamma = \frac{1}{2}\frac{1}{2J+1}\frac{1}{(2\pi)^3}\frac{1}{32 m_{\chi_{cJ}}^3} \overline{\left|\mathcal{M}_{\mathrm{tot}}\right|^2} dm_{\eta\eta}^2 dm_{\eta\eta^\prime}^2\,.
	\end{align}
	The factor $1/2$ is due to the presence of two identical $\eta$ mesons in the final state; the factor $1/(2J+1)$ originates from the average over the spin states of the initial $\chi_{cJ}$ meson; the overline indicates the summation over the polarization states of the initial $\chi_{cJ}$ meson. The $\eta\eta$ and $\eta\eta^\prime$ invariant mass squares are denoted as $m_{\eta\eta}^2 = (p_1+p_2)^2$ and $m_{\eta\eta^\prime}^2 = (p_2+p_3)^2$, respectively, where $p_{1(2,3)}$ are the four-momenta of the final $\eta$ and $\eta^\prime$ mesons, as indicated in Fig.~\ref{fig:feyndiagschic2etas}.
  
\section{Numerical Results and Discussion}\label{sec:results}

\subsection{Estimate of coupling $g_{f_0}$}

    Under the SU(4) symmetry, the coupling $g_{f_0}$ in Eq.~\eqref{eq:gf0} could be extracted from the experimental data on the scalar $f_0(1500)$ decaying into the light pseudoscalar mesons, such as $\pi\pi$ or $K\bar{K}$. Using $\mathcal{B}[f_0(1500)\to K\bar{K}]=(8.5\pm 1.0)\%$ \cite{ParticleDataGroup:2024cfk} yields $g_{f_0} = (1.37\pm 0.22)$, while $\mathcal{B}[f_0(1500)\to \pi\pi] = (34.5\pm 2.2)\%$ \cite{ParticleDataGroup:2024cfk} gives $g_{f_0} = (10.1\pm 1.6)$. Therefore, the average value of $g_{f_0} = (5.7\pm 0.8)$ could serve as an reasonable estimation of the coupling under the SU(4) symmetry.

    Another rough estimate of the $g_{f_0}$ can be modeled on the widely used couplings of the scalar mesons $f_0(500)$ and $f_0(980)$ to the charmed mesons \cite{Meng:2008dd,Bai:2022cfz,Chen:2011qx,Chen:2015bma}, which are $3.73/\sqrt{6}$ and $3.73/\sqrt{3}$, respectively. Taking into account the effect of the scalar meson masses, $g_{f_0}$ could be estimated to be $g_{f_0} = 3.73/ \sqrt{2} = 2.64$, which is more than two times smaller than the value of $g_{f_0} = (5.7\pm 0.8)$ from the SU(4) symmetry.

\begin{figure}
		\centering
		\includegraphics[width=0.94\linewidth]{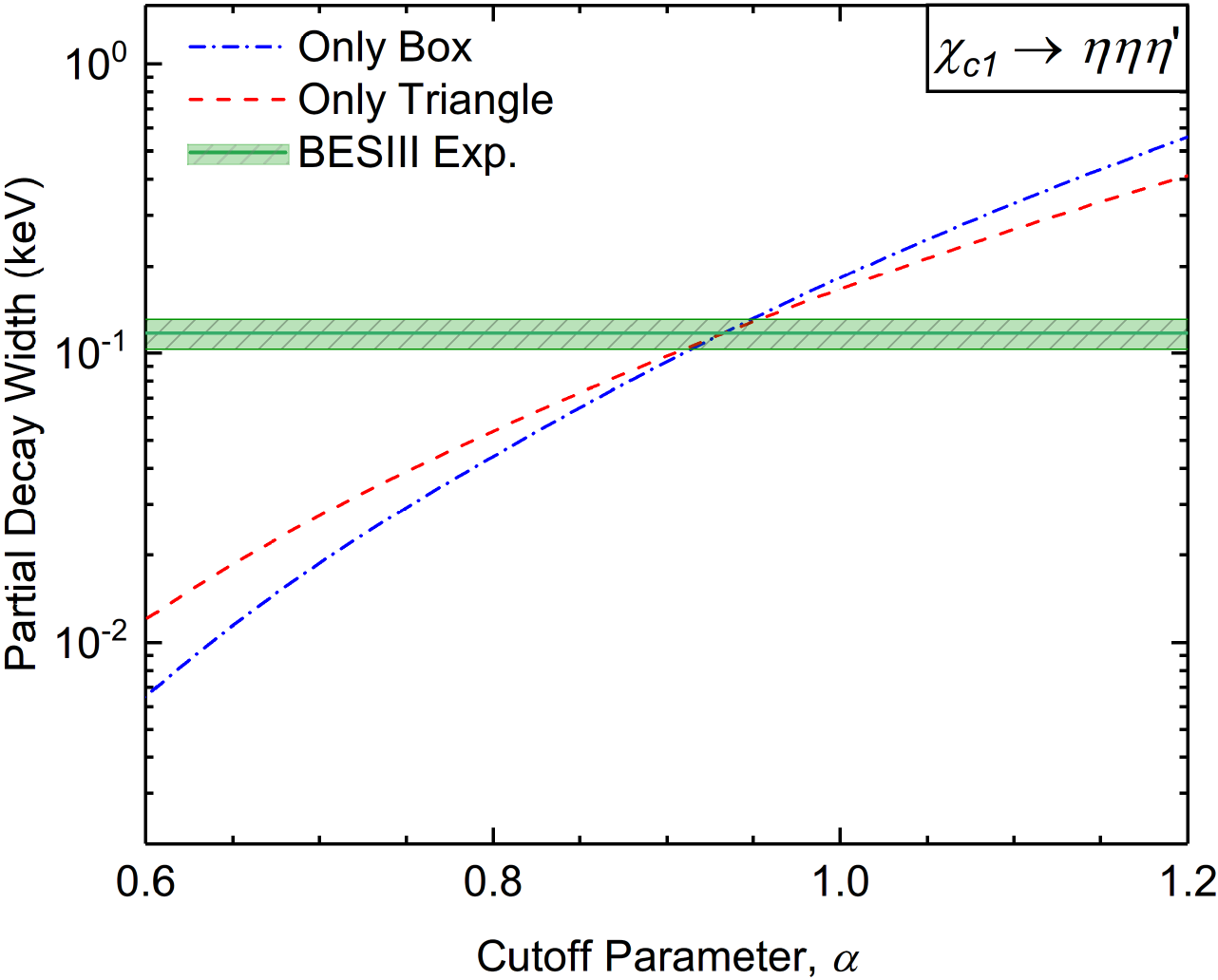}
		\caption{Partial decay width of the $\chi_{c1} \to \eta \eta \eta^\prime$ process due to the only box or triangle loops for different values of the cutoff parameter $\alpha$. The triangle loop contributions are calculated with the coupling $g_{f_0} = 6.87$.
		The experimental decay width measured by the BESIII Collaboration is $(0.117\pm 0.014)~\mathrm{keV}$ \cite{BESIII:2025izw}, as indicated by the band. }
		\label{fig:widthChic1DtmGf0}
	\end{figure}

    The realistic estimate of the coupling $g_{f_0}$ can be obtained by considering the BESIII partial wave analysis of the $\eta \eta^\prime$ invariant mass spectrum in the $\chi_{c1} \to \eta \eta \eta^\prime$ decay, which indicates that the structure in the $\eta \eta^\prime$ mass spectrum is mainly attributed to the $f_0(1500)$~\cite{BESIII:2025izw}. This means that the $f_0(1500)$ contribution to the $\chi_{c1} \to \eta \eta \eta^\prime$ decay is significant. Accordingly, we could first fit the experimental branching fraction of $\chi_{c1} \to \eta\eta\eta^\prime$ by considering only the box loop contributions, with the cutoff $\alpha$ in the form factor as the only free parameter (all couplings in the box diagrams are well known). At the fitted value of $\alpha$ (around $\alpha =0.93$, see Fig.~\ref{fig:widthChic1DtmGf0}), we then fit the triangle loop contributions involving only the intermediate $f_0(1500)$ through the same experimental branching fraction to determine $g_{f_0}$. This procedure yields $g_{f_0} \approx 6.87$, which is close to the SU(4) value of $(5.7\pm 0.8)$. Consequently, $g_{f_0}=6.87$ is adopted throughout unless otherwise stated.
    

\subsection{$\chi_{c1}\to\eta\eta\eta^\prime$}

\begin{figure}
		\centering
		\includegraphics[width=0.98\linewidth]{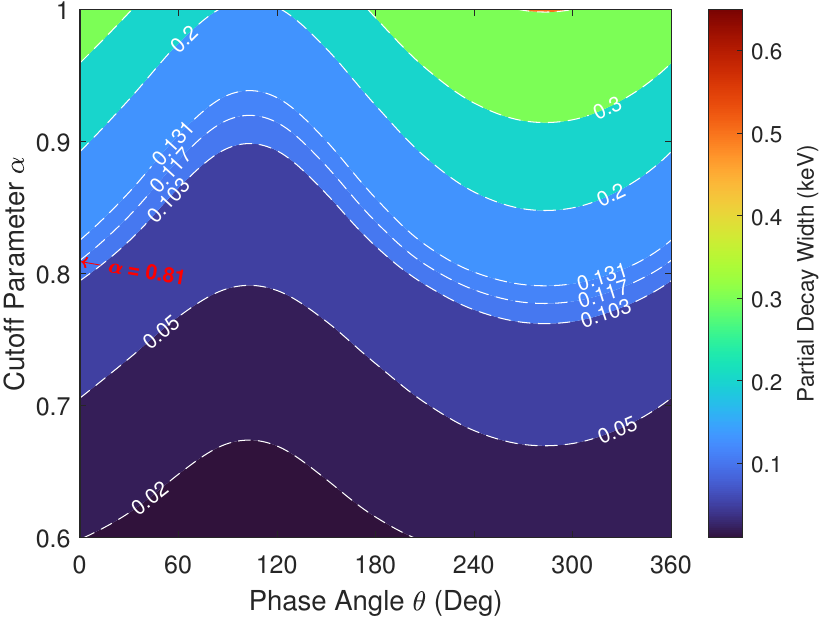}
		\caption{Partial decay width of the $\chi_{c1} \to \eta \eta \eta^\prime$ process for different values of the cutoff parameter $\alpha$ and the relative phase angle $\theta$. The experimental decay width measured by the BESIII Collaboration is $(0.117\pm 0.014)~\mathrm{keV}$ \cite{BESIII:2025izw}, as indicated by the three densely spaced lines representing the central value and uncertainties. }
		\label{fig:widthchic1vsalphaangle}
	\end{figure}

	With the above formalism, we can conduct the numerical calculations of the decay processes $\chi_{cJ} \to \eta \eta \eta^\prime$. First, we focus on the process $\chi_{c1} \to \eta \eta \eta^\prime$ since BESIII gave more information about this process, such as the partial decay width and the invariant mass distributions of the final meson pairs \cite{BESIII:2025izw}.
	For the process $\chi_{c1} \to \eta \eta \eta^\prime$, there are two free parameters in our calculations, i.e., the cutoff parameter $\alpha$ in the form factor and the relative phase angle $\theta$ between the box and triangle loop amplitudes. Figure \ref{fig:widthchic1vsalphaangle} shows the partial decay width of the $\chi_{c1} \to \eta \eta \eta^\prime$ for different values of $\alpha$ and $\theta$. It can be seen that at given value of $\theta$, the partial decay width increases with the value of $\alpha$. As expected, at fixed value of $\alpha$, the partial decay width varies periodically with $\theta$ due to the interference between the box and triangle loop amplitudes [see Eq. \eqref{eq:amptot}]. 

\begin{figure}
		\centering
		\includegraphics[width=0.9\linewidth]{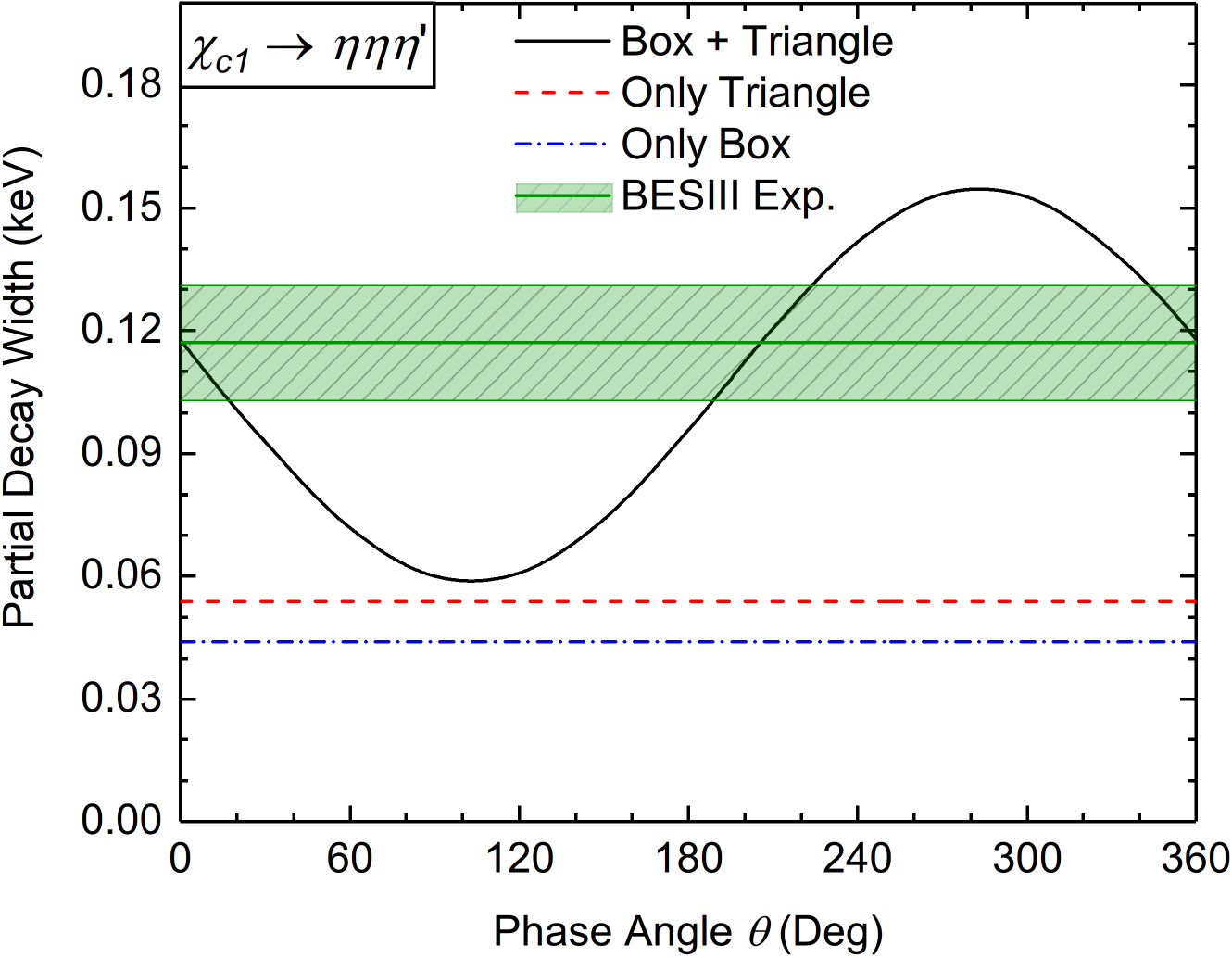}
		\caption{Partial decay widths of the $\chi_{c1}\to \eta \eta \eta^\prime$ process as a function of the relative phase angle $\theta$ defined in Eq. \eqref{eq:amptot}. The calculations are performed at $\alpha = 0.81$. The black solid line represents the total decay width of the $\chi_{c1}\to \eta \eta \eta^\prime$ process with both box and triangle loop contributions. The horizontal dash-dotted line indicates only the box loop contribution, while the horizontal dashed line is only the triangle loop contribution. The band represents the experimental decay width measured by the BESIII Collaboration \cite{BESIII:2025izw}. }
		\label{fig:widthofboxvstri}
	\end{figure}

	The decay width of the $\chi_{c1}\to \eta \eta \eta^\prime$ measured by the BESIII Collaboration is $(0.117\pm 0.014)~\mathrm{keV}$, corresponding to the branching fraction of $(1.39\pm 0.16)\times 10^{-4}$ \cite{BESIII:2025izw}. Our calculations can reproduce this width with values of $\alpha$ in the range of $0.76\leq \alpha \leq 0.94$. It is noticed that the predicted decay width at a given value of $\alpha$ varies significantly with the relative phase angle $\theta$. This is due to the interference between the triangle and box loop contributions. In Fig. \ref{fig:widthofboxvstri}, we show the partial decay widths of the $\chi_{c1}\to \eta \eta \eta^\prime$ process as a function of the relative phase angle $\theta$ at $\alpha = 0.81$, along with the separate contributions from the triangle and box diagrams. The experimental measurement is also indicated for comparison. From Fig. \ref{fig:widthofboxvstri}, we can see that partial decay width contributed from the triangle loops is about 0.054 keV, a little greater than that (about 0.044 keV) due to the box loop contribution. The role of the interference between the triangle and box loop contributions in influencing the invariant mass distributions of the final meson pairs will be discussed below.

\begin{figure*}[htbp]
		\centering
		\includegraphics[width=0.96\linewidth]{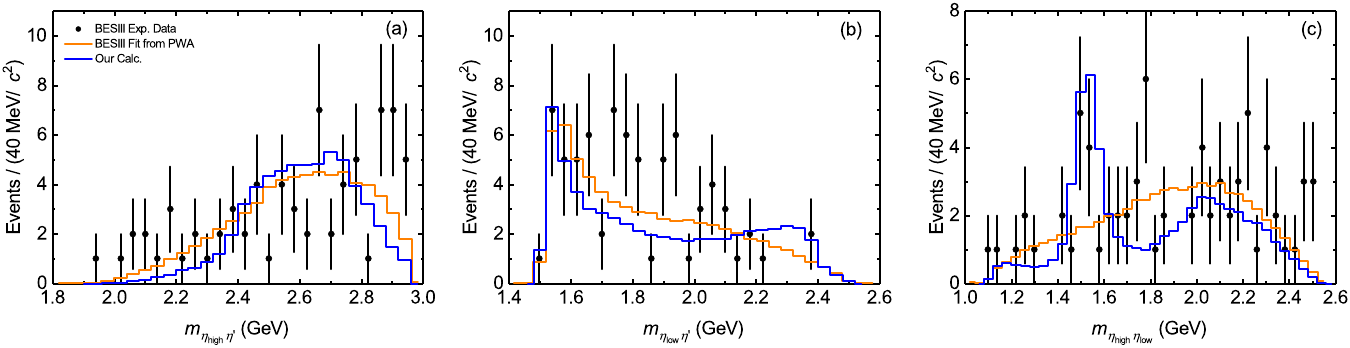}\\
		\includegraphics[width=0.96\linewidth]{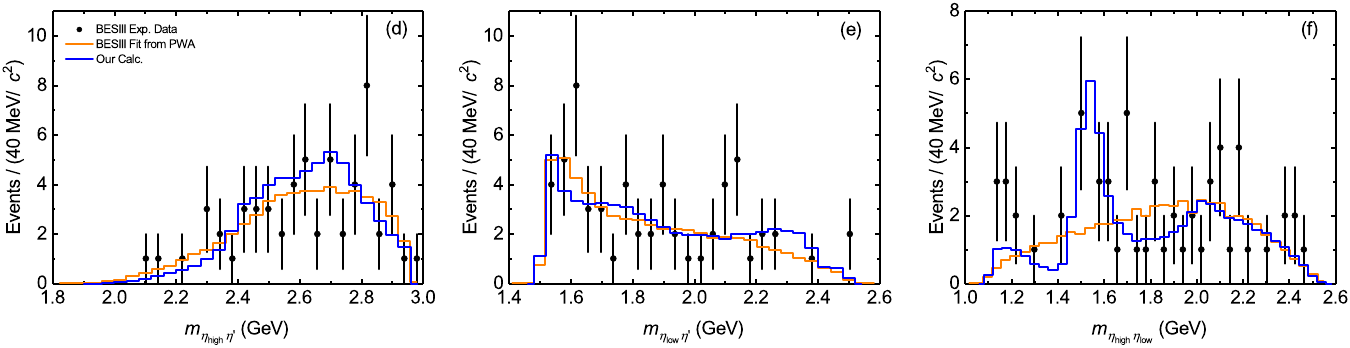}
		\caption{Invariant mass distributions of the $\eta_{\mathrm{high}}\eta^\prime$ [(a) and (d)], $\eta_{\mathrm{low}}\eta^\prime$ [(b) and (e)] and $\eta_{\mathrm{high}}\eta_{\mathrm{low}}$ [(c) and (f)] pairs in the $\chi_{c1}\to \eta \eta \eta^\prime$ process. The $\eta_{\mathrm{low}}$ and $\eta_{\mathrm{high}}$ denote the lower and higher energy $\eta$ mesons in the final state, respectively. For the BESIII data in the panels (a)-(c), the $\eta^\prime$ meson is reconstructed via the $\gamma\pi^+\pi^-$, while in the panels (d)-(f), the $\eta^\prime$ meson is reconstructed with the $\eta\pi^+\pi^-$. Our calculated results (blue line) are compared with the BESIII measurements \cite{BESIII:2025izw} (black points with error bars). The partial wave analysis by the BESIII Collaboration is also shown as the orange line. Our calculations for (a)--(c) are performed at $\theta = 210^\circ$ and $\alpha=0.81$, while those for (d)--(f) are conducted at $\theta = 150^\circ$ and $\alpha=0.88$. These results are based on Monte Carlo simulations with 59000 generated events, statistically normalized to the experimental yield of 59 events.}
		\label{fig:invariantMassDistributionOfChic1Comparison}
	\end{figure*}

	Our calculations show that the invariant mass distributions of the final states are not very sensitive to cutoff parameter $\alpha$, but, due to the interference effect between the box and triangle loop contributions, they depend clearly on the relative phase angle $\theta$. To constrain the relative phase angle $\theta$, we performed a comparison of the theoretical results with the BESIII measurements of the $\eta\eta$ and $\eta\eta^\prime$ invariant mass distributions of the $\chi_{c1}\to \eta \eta \eta^\prime$ process. The comparison indicates that the better agreement between our theoretical results and the BESIII measurements can be achieved around $\theta = 150^\circ \sim 210^\circ$. The results calculated at $\theta = 180^\circ$ and $\theta = 210^\circ$ are shown in Fig. \ref{fig:invariantMassDistributionOfChic1Comparison}, where our calculated results (blue line) are compared with the BESIII measurements (black points with error bars) \cite{BESIII:2025izw}. The partial wave analysis (PWA) by the BESIII Collaboration is also shown as the orange line. From Fig. \ref{fig:invariantMassDistributionOfChic1Comparison}, we can see that our calculated results are overall consistent with the BESIII measurements, which indicates that the charmed meson loop mechanisms can reasonably describe the decay processes $\chi_{c1} \to \eta \eta \eta^\prime$. However, there exist some discrepancies between our theoretical results and the BESIII PWA analysis, especially for the $\eta_{\mathrm{high}}\eta_{\mathrm{low}}$ invariant mass distribution. These discrepancies are straightforward in view of the fact that the BESIII PWA analyses were performed by considering the $f_0(1500)$ resonance and the $S$-wave nonresonant contribution \cite{BESIII:2025izw}, while our calculations consider the contributions from the charmed meson loops. Compared to the BESIII PWA analysis, our predicted invariant mass distributions could reproduce the possible peak structure around 1.5 GeV in the $\eta_{\mathrm{high}}\eta_{\mathrm{low}}$ invariant mass spectrum.

	The overall consistency between our theoretical results and the BESIII measurements also helps explain the non-observation of the $\eta_1(1855)$ signal in this channel under the present experimental statistics. According to the foregoing discussions and the results shown in Figs. \ref{fig:widthChic1DtmGf0}-\ref{fig:invariantMassDistributionOfChic1Comparison}, the partial decay width of the $\chi_{c1}\to \eta \eta \eta^\prime$ process is dominated by the triangle and box loop contributions. If the present loop mechanism is the true underlying physical situation for the $\chi_{c1}\to \eta \eta \eta^\prime$ process, then the contributions of other possible intermediate resonances, such as the $\eta_1(1855)$, would be at most comparable to that of the $f_0(1500)$. According to the current prediction, at $\alpha =0.88$, the partial decay width only due to the triangle loop contributions is about 0.088 keV. Therefore, the branching fraction $\mathcal{B}[\chi_{c1}\to f_0(1500)\eta^{(\prime)} \to \eta\eta\eta^\prime]$ is around $1.05 \times 10^{-4}$. Theoretically, we could take the $f_0(1500)$ branching fraction as a conservative upper estimate, namely $\mathcal{B}[\chi_{c1}\to \eta_1(1855)\eta^{(\prime)} \to \eta\eta\eta^\prime] \leq 1.05 \times10^{-4}$, in agreement with the upper limit of $\mathcal{B}[\chi_{c1}\to \eta_1(1855)\eta] \cdot \mathcal{B}[\eta_1(1855)\to \eta\eta^\prime] < 9.79\times 10^{-5}$ given by the BESIII Collaboration \cite{BESIII:2025izw}.

	\subsection{$\chi_{c2}\to\eta\eta\eta^\prime$}

	\begin{figure}[htbp]
		\centering
		\includegraphics[width=0.9\linewidth]{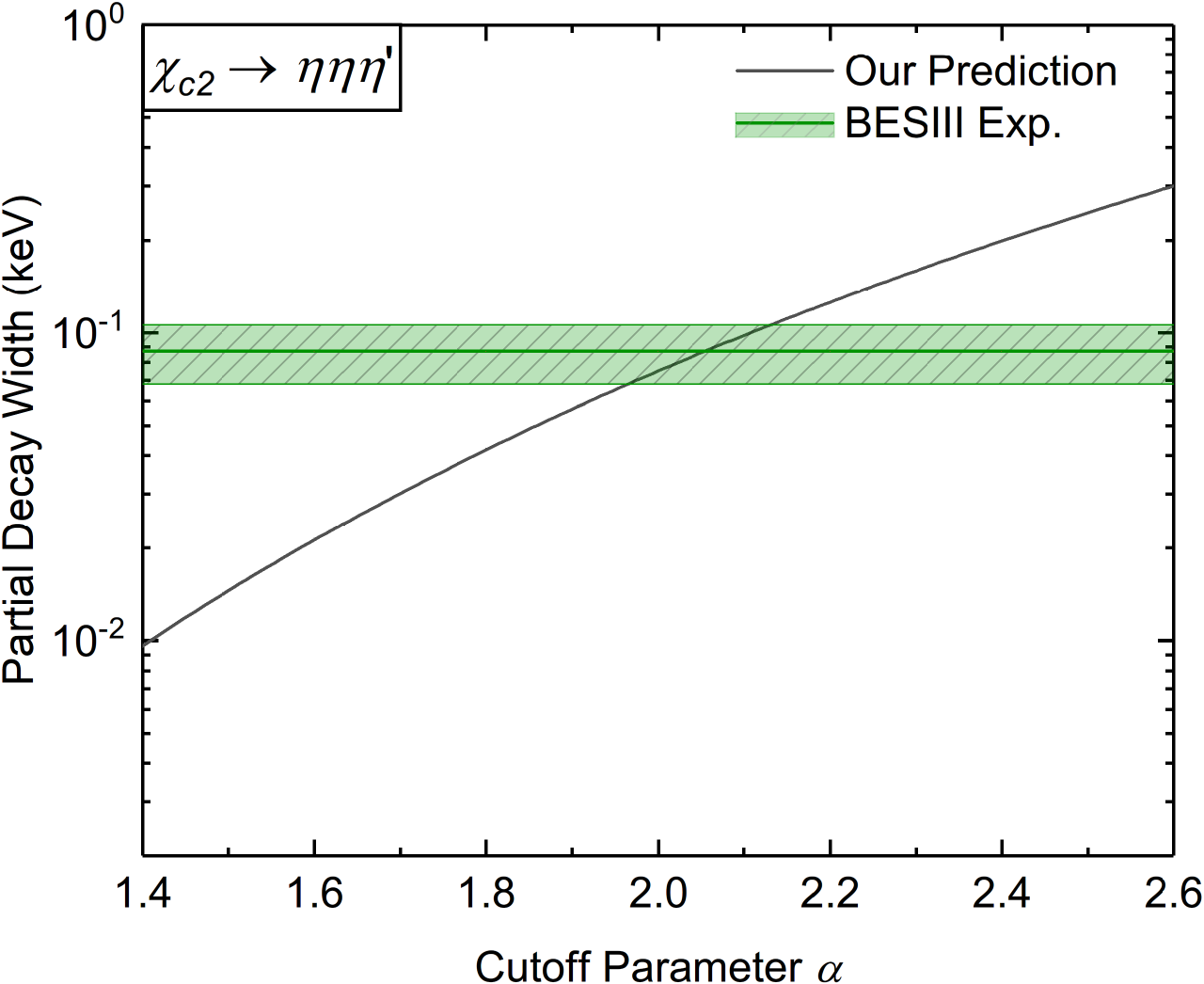}
		\caption{Partial decay width of the $\chi_{c2} \to \eta \eta \eta^\prime$ process for different values of the cutoff parameter $\alpha$. The band indicates the experimental decay width of $(0.087\pm 0.019)~\mathrm{keV}$ measured by the BESIII Collaboration \cite{BESIII:2025izw}.}
		\label{fig:widthchic2vsalpha}
	\end{figure}

	In the case of $\chi_{c2}\to\eta\eta\eta^\prime$, only the box loop amplitude contributes under the current Lagrangian framework. The measured decay width of the $\chi_{c2} \to \eta \eta \eta^\prime$ by the BESIII Collaboration is $(0.087\pm 0.019)~\mathrm{keV}$, corresponding to a branching fraction of $(4.42\pm 0.93)\times 10^{-5}$ \cite{BESIII:2025izw}. In Fig. \ref{fig:widthchic2vsalpha}, the partial decay width of the $\chi_{c2} \to \eta \eta \eta^\prime$ process is shown for different values of $\alpha$. It can be seen that our calculations can reproduce the BESIII measurement around $\alpha = 2.0$. 

\begin{figure}
		\centering
		\includegraphics[width=0.98\linewidth]{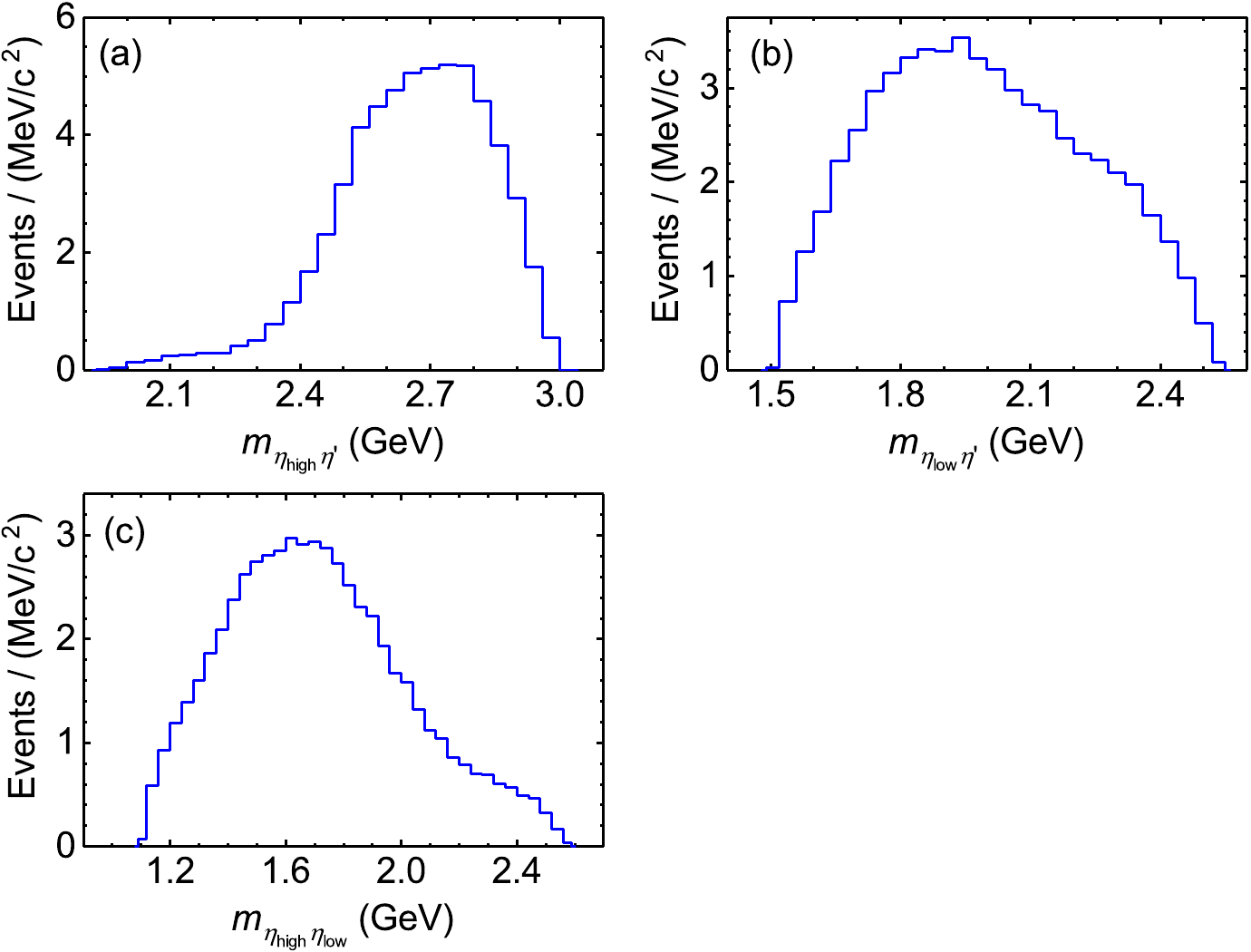}
		\caption{The predicted invariant mass distributions of the $\eta_{\mathrm{high}}\eta^\prime$ (a), $\eta_{\mathrm{low}}\eta^\prime$ (b), and $\eta_{\mathrm{high}}\eta_{\mathrm{low}}$ (c) pairs in the $\chi_{c2}\to\eta\eta\eta^\prime$ process. The $\eta_{\mathrm{low}}$ and $\eta_{\mathrm{high}}$ represent the lower and higher energy $\eta$ mesons in the final states, respectively. The distribution curves are obtained at $\alpha = 2.0$ based on Monte Carlo simulations with 59000 generated events. }
		\label{fig:invariantMassDistributionOfChic2}
	\end{figure}

	In Fig. \ref{fig:invariantMassDistributionOfChic2}, we show the theoretical invariant mass distributions of the $\eta_{\mathrm{high}}\eta^\prime$, $\eta_{\mathrm{low}}\eta^\prime$ and $\eta_{\mathrm{high}}\eta_{\mathrm{low}}$ pairs in the $\chi_{c2}\to\eta\eta\eta^\prime$ process. They can be used to compare with the future BESIII measurements, which would be helpful to check the validity of the box loop mechanisms for the $\chi_{c2}\to\eta\eta\eta^\prime$ process. It should be pointed out that some possible intermediate resonances might contribute to the $\chi_{c2}\to\eta\eta\eta^\prime$ process. However, similar to the case of $\chi_{c1}\to\eta\eta\eta^\prime$, the box loop mechanism is also expected to be the dominant mechanism for the $\chi_{c2}\to\eta\eta\eta^\prime$ process.

\section{Summary}\label{sec:summary}

	Motivated by the first measurements of the $\chi_{cJ}\to\eta\eta\eta^\prime$ by the BESIII collaboration \cite{BESIII:2025izw}, we have investigated these processes using an effective Lagrangian approach. The $\chi_{cJ}\to\eta\eta\eta^\prime$ decays are assumed to occur via charmed meson loops. For the $\chi_{c1} \to \eta \eta \eta^\prime$ process, the box and triangle loops are considered, while only the box loops are considered for the $\chi_{c2} \to \eta \eta \eta^\prime$ process. 
	
	The experimental decay widths of the $\chi_{cJ} \to \eta \eta \eta^\prime$ processes can be reproduced well by taking reasonable values of the cutoff parameter in the form factor. The invariant mass distributions of the final meson pairs in the $\chi_{c1} \to \eta \eta \eta^\prime$ process that are predicted using the box and triangle loops are overall consistent with the BESIII measurements. Our calculations show that the $\chi_{c1} \to \eta \eta \eta^\prime$ process is nearly saturated by the triangle and box loop contribution. This implies that the other possible intermediate resonances, e.g. the $\eta_1(1855)$, if it could be produced in these processes, would contribute no more than the $f_0(1500)$ does. Under the assumption that the present loop mechanism is the true underlying physical situation for the $\chi_{c1}\to \eta \eta \eta^\prime$ process, the branching fraction $\mathcal{B}[\chi_{c1}\to \eta_1(1855)\eta^{(\prime)} \to \eta\eta\eta^\prime]$ is speculated to be at most $ 1.05\times 10^{-4}$. Finally, we predicted the invariant mass distributions of the final meson pairs in the $\chi_{c2} \to \eta \eta \eta^\prime$ process, which can be directly used to compare with future BESIII measurements. The comparison is helpful to check the validity of the box loop mechanisms for the $\chi_{c2}\to\eta\eta\eta^\prime$ process. The present work might be extended to the $\chi_{cJ}$ decays into other three light mesons, for instance the $\chi_{cJ} \to\phi\phi\eta^{(\prime)}$ \cite{BESIII:2025wxa}.

\begin{acknowledgments}\label{sec:acknowledgements}
    This work is supported by the National Natural Science Foundation of China under Grant Nos. 12475081, 12405093, 12575094, 12435007, and 12361141819; by the Natural Science Foundation of Shandong Province under Grant No. ZR2025MS04; by Taishan Scholar Project of Shandong Province; and by National Key Research and Development Program under Grant Nos. 2023YFA1606703 and 2024YFA1610504.	
\end{acknowledgments}
		

	\bibliography{particlePhysClean.bib}

@misc{Wan:2025sae,
    author = "Wan, Bing-Dong and Zhang, Yan and Zhang, Jun-Hao and Yuan, Ming-Yang",
    title = "{Hidden-charm and -bottom tetraquark states with $J^{PC}=1^{-+}$ via QCD sum rules}",
    eprint = "2512.03800",
    archivePrefix = "arXiv",
    primaryClass = "hep-ph",
    month = "12",
    year = "2025"
}

@misc{BESIII:2025wxa,
    author = "Ablikim, Medina and others",
    collaboration = "BESIII",
    title = "{Measurements of the branching fractions of $\chi_{cJ}\to \phi\phi\eta, \phi\phi\eta^{\prime}$ and $\phi K^+K^-\eta$}",
    eprint = "2512.14369",
    archivePrefix = "arXiv",
    primaryClass = "hep-ex",
    reportNumber = "BAM-00909",
    month = "12",
    year = "2025"
}

@article{COMPASS:2009xrl,
    author = "Alekseev, M. and others",
    collaboration = "COMPASS",
    title = "{Observation of a $J^{PC} = 1^{-+}$ exotic resonance in diffractive dissociation of 190 GeV/$c$ $\pi^-$ into $\pi^- \pi^- \pi^+$}",
    eprint = "0910.5842",
    archivePrefix = "arXiv",
    primaryClass = "hep-ex",
    reportNumber = "CERN-PH-EP-2009-018",
    doi = "10.1103/PhysRevLett.104.241803",
    journal = "Phys. Rev. Lett.",
    volume = "104",
    pages = "241803",
    year = "2010"
}

@article{Chung:2002pu,
    author = "Chung, S. U. and others",
    title = "{Exotic and $q\bar{q}$ resonances in the $\pi^+ \pi^- \pi^-$ system produced in $\pi^- p$ collisions at 18 GeV/$c$}",
    reportNumber = "JLAB-PHY-01-34",
    doi = "10.1103/PhysRevD.65.072001",
    journal = "Phys. Rev. D",
    volume = "65",
    pages = "072001",
    year = "2002"
}

@misc{BESIII:2025izw,
    author = "Ablikim, Medina and others",
    collaboration = "BESIII",
    title = "{Search for $\eta_{1}(1855)$ in $\chi_{cJ}\to\eta\eta\eta^{\prime}$ decays}",
    eprint = "2504.19087",
    archivePrefix = "arXiv",
    primaryClass = "hep-ex",
    month = "4",
    year = "2025"
}

@article{Qiu:2022ktc,
  title = "{Towards the Establishment of the Light $J^{PC}=1^{-(+)}$ Hybrid Nonet}",
  author = {Qiu, Lin and Zhao, Qiang},
  year = 2022,
  journal = {Chin. Phys. C},
  volume = {46},
  number = {5},
  eprint = {2202.00904},
  primaryclass = {hep-ph},
  pages = {051001},
  issn = {1674-1137, 2058-6132},
  doi = {10.1088/1674-1137/ac567e},
  urldate = {2025-10-11},
  archiveprefix = {arXiv},
}

@article{Chen:2011qx,
  title = "{Dipion invariant mass distribution of the anomalous $\ensuremath{\Upsilon}(1S){\ensuremath{\pi}}^{+}{\ensuremath{\pi}}^{\ensuremath{-}}$ and $\ensuremath{\Upsilon}(2S){\ensuremath{\pi}}^{+}{\ensuremath{\pi}}^{\ensuremath{-}}$ production near the peak of $\ensuremath{\Upsilon}(10860)$}",
  author = {Chen, Dian-Yong and He, Jun and Li, Xue-Qian and Liu, Xiang},
  year = 2011,
  journal = {Phys. Rev. D},
  volume = {84},
  number = {7},
  eprint = {1105.1672},
  primaryclass = {hep-ph},
  pages = {074006},
  publisher = {American Physical Society},
  doi = {10.1103/PhysRevD.84.074006},
  urldate = {2023-10-25},
  archiveprefix = {arXiv},
}

@article{WA102:1999hsn,
    author = "Barberis, D and others",
    collaboration = "WA102",
    title = "{A Study of the $\eta\eta^\prime$ and $\eta^\prime\eta^\prime$ channels produced in central p p interactions at 450-GeV/c}",
    eprint = "hep-ex/9911041",
    archivePrefix = "arXiv",
    doi = "10.1016/S0370-2693(99)01392-1",
    journal = "Phys. Lett. B",
    volume = "471",
    pages = "429--434",
    year = "2000"
}

@article{Zaitsev:2000rc,
    author = "Zaitsev, A.",
    collaboration = "VES",
    title = "{Study of exotic resonances in diffractive reactions}",
    doi = "10.1016/S0375-9474(00)00238-4",
    journal = "Nucl. Phys. A",
    volume = "675",
    pages = "155C--160C",
    year = "2000"
}

@article{Li:2021jjt,
  title = {Hidden-bottom hadronic decays of $\mathrm{\ensuremath{\Upsilon}}(10753)$ with a ${{\eta}}^{({\prime})}$ or $\ensuremath{\omega}$ emission},
  author = {Li, Yu-Shuai and Bai, Zi-Yue and Huang, Qi and Liu, Xiang},
  year = 2021,
  month = aug,
  journal = {Phys. Rev. D},
  volume = {104},
  number = {3},
  eprint = {2106.14123},
  primaryclass = {hep-ph},
  pages = {034036},
  publisher = {American Physical Society},
  doi = {10.1103/PhysRevD.104.034036},
  urldate = {2023-04-28},
  archiveprefix = {arXiv},
}

@article{Qiao:2023dhf,
    author = "Qiao, Cong-Feng and Wan, Bing-Dong",
    title = "{The natures of recently observed states ${\eta}_1(1855)$ and $X(2600)$}",
    doi = "10.1016/j.nuclphysbps.2023.01.017",
    journal = "Nucl. Part. Phys. Proc.",
    volume = "324-329",
    pages = "76--79",
    year = "2023"
}

@article{COMPASS:2018uzl,
    author = "Aghasyan, M. and others",
    collaboration = "COMPASS",
    title = "{Light isovector resonances in $\pi^- p \to \pi^-\pi^-\pi^+ p$ at 190 GeV/${\it c}$}",
    eprint = "1802.05913",
    archivePrefix = "arXiv",
    primaryClass = "hep-ex",
    reportNumber = "CERN-EP-2018-021",
    doi = "10.1103/PhysRevD.98.092003",
    journal = "Phys. Rev. D",
    volume = "98",
    number = "9",
    pages = "092003",
    year = "2018"
}

@article{Kopf:2020yoa,
    author = {Kopf, B. and Albrecht, M. and Koch, H. and K{\"u}{\ss}ner, M. and Pychy, J. and Qin, X. and Wiedner, U.},
    title = "{Investigation of the lightest hybrid meson candidate with a coupled-channel analysis of ${{\bar{p}}p}$-, $\pi ^- p$- and ${\pi \pi }$-Data}",
    eprint = "2008.11566",
    archivePrefix = "arXiv",
    primaryClass = "hep-ph",
    doi = "10.1140/epjc/s10052-021-09821-2",
    journal = "Eur. Phys. J. C",
    volume = "81",
    number = "12",
    pages = "1056",
    year = "2021"
}

@article{Liu:2025sjz,
  title = "{Hadronic Decays of Possible Pseudoscalar $P$-wave $D\bar{D}^*/\bar{D}D^*$ Molecular State}",
  author = {Liu, Shi-Dong and Wu, Qi and Li, Gang},
  year = 2025,
  month = oct,
  journal = {Phys. Rev. D},
  volume = {112},
  number = {7},
  eprint = {2506.18273},
  primaryclass = {hep-ph},
  pages = {074002},
  doi = {10.1103/t3q7-fml8},
  archiveprefix = {arXiv},
}

@article{Chen:2023ukh,
  title = {Constructing the $J^{P(C)} =1^{-(+)}$ Light Flavor Hybrid Nonet with the Newly Observed ${\eta_1(1855)}$},
  author = {Chen, Bing and Luo, Si-Qiang and Liu, Xiang},
  year = 2023,
  month = sep,
  journal = {Phys. Rev. D},
  volume = {108},
  number = {5},
  eprint = {2302.06785},
  primaryclass = {hep-ph},
  pages = {054034},
  issn = {2470-0010, 2470-0029},
  doi = {10.1103/PhysRevD.108.054034},
  urldate = {2025-10-11},
  archiveprefix = {arXiv},
}

@article{Meng:2008dd,
  title = "{Peak shifts due to ${B}^{(*)}\ensuremath{-}{\overline{B}}^{(*)}$ rescattering in $\ensuremath{\Upsilon}(5S)$ dipion transitions}",
  author = {Meng, Ce and Chao, Kuang-Ta},
  year = 2008,
  journal = {Phys. Rev. D},
  volume = {78},
  number = {3},
  eprint = {0805.0143},
  primaryclass = {hep-ph},
  pages = {034022},
  publisher = {American Physical Society},
  doi = {10.1103/PhysRevD.78.034022},
  urldate = {2023-11-21},
  archiveprefix = {arXiv},
}

@article{Wang:2010iq,
  title = {Study of $\eta_c$ and $\eta_c^\prime$ decays into vector meson pairs},
  author = {Wang, Qian and Liu, Xiao-Hai and Zhao, Qiang},
  year = 2010,
  month = oct,
  journal = {arXiv:1010.1343 [hep-ph]},
  eprint = {1010.1343},
  primaryclass = {hep-ph},
  archiveprefix = {arXiv},
}

@article{Khokhlov:2000tk,
    author = "Khokhlov, Yu. A.",
    editor = "Faldt, G. and Hoistad, B. and Kullander, S.",
    collaboration = "VES",
    title = "{Study of $X(1600)~1^{-+}$ hybrid}",
    doi = "10.1016/S0375-9474(99)00663-6",
    journal = "Nucl. Phys. A",
    volume = "663",
    pages = "596--599",
    year = "2000"
}

@article{Yan:2023vbh,
  title = {On the $\eta_1(1855)$, $\pi_1(1400)$ and $\pi_1(1600)$ as {{Dynamically Generated States}} and {{Their SU}}(3) {{Partners}}},
  author = {Yan, Mao-Jun and Dias, Jorgivan M. and Guevara, Adolfo and Guo, Feng-Kun and Zou, Bing-Song},
  year = 2023,
  month = feb,
  journal = {Universe},
  volume = {9},
  number = {2},
  eprint = {2301.04432},
  primaryclass = {hep-ph},
  pages = {109},
  doi = {10.3390/universe9020109},
  archiveprefix = {arXiv},
}

@article{Liang:2024lon,
  title = {Decay Properties of Light $1^{-+}$ Hybrids},
  author = {Liang, Juzheng and Chen, Siyang and Chen, Ying and Shi, Chunjiang and Sun, Wei},
  year = 2025,
  month = feb,
  journal = {Sci. China Phys. Mech. Astron.},
  volume = {68},
  number = {5},
  eprint = {2409.14410},
  primaryclass = {hep-lat},
  pages = {251011},
  doi = {10.1007/s11433-024-2588-0},
  archiveprefix = {arXiv},
}

@article{CLEO:2011upl,
    author = "Adams, G. S. and others",
    collaboration = "CLEO",
    title = "{Amplitude analyses of the decays $\chi_{c1} \to \eta \pi^+ \pi^-$ and $\chi_{c1} \to \eta' \pi^+ \pi^-$}",
    eprint = "1109.5843",
    archivePrefix = "arXiv",
    primaryClass = "hep-ex",
    reportNumber = "CLNS-11-2080, CLEO-11-06",
    doi = "10.1103/PhysRevD.84.112009",
    journal = "Phys. Rev. D",
    volume = "84",
    pages = "112009",
    year = "2011"
}

@article{Zhang:2025gmm,
    author = "Zhang, Xiao-Yu and Shi, Pan-Pan and Guo, Feng-Kun",
    title = "{Production of $1^{-+}$ exotic charmonium-like states in electron-positron collisions}",
    eprint = "2503.06259",
    archivePrefix = "arXiv",
    primaryClass = "hep-ph",
    doi = "10.1016/j.physletb.2025.139603",
    journal = "Phys. Lett. B",
    volume = "867",
    pages = "139603",
    year = "2025"
}

@article{Shi:2023sdy,
  title = {Decays of ${1}^{\ensuremath{-}+}$ charmoniumlike hybrid using lattice QCD},
  author = {Shi, Chunjiang and Chen, Ying and Gong, Ming and Jiang, Xiangyu and Liu, Zhaofeng and Sun, Wei},
  year = 2024,
  month = may,
  journal = {Phys. Rev. D},
  volume = {109},
  number = {9},
  eprint = {2306.12884},
  primaryclass = {hep-lat},
  pages = {094513},
  doi = {10.1103/PhysRevD.109.094513},
  urldate = {2025-04-21},
  archiveprefix = {arXiv},
}

@misc{Xu:2025epx,
    author = "Xu, Kai and Zhao, Zheng and Tagsinsit, Nattapat and Kaewsnod, Attaphon and Limphirat, Ayut and Herold, Christoph and Yan, Yupeng",
    title = "{Systematic study of exotic $1^{-+}$ tetraquark spectroscopy}",
    eprint = "2511.22111",
    archivePrefix = "arXiv",
    primaryClass = "hep-ph",
    month = "11",
    year = "2025"
}

@article{Wan:2022xkx,
  title = {Possible Structure of the Newly Found Exotic State $\eta_1(1855)$},
  author = {Wan, Bing-Dong and Zhang, Sheng-Qi and Qiao, Cong-Feng},
  year = 2022,
  month = oct,
  journal = {Phys. Rev. D},
  volume = {106},
  number = {7},
  eprint = {2203.14014},
  primaryclass = {hep-ph},
  pages = {074003},
  issn = {2470-0010, 2470-0029},
  doi = {10.1103/PhysRevD.106.074003},
  urldate = {2025-10-11},
  archiveprefix = {arXiv}
}

@article{Cheng:2004ru,
  title = "{Final state interactions in hadronic $B$ decays}",
  author = {Cheng, Hai-Yang and Chua, Chun-Khiang and Soni, Amarjit},
  year = 2005,
  journal = {Phys. Rev. D},
  volume = {71},
  number = {1},
  eprint = {hep-ph/0409317},
  pages = {014030},
  publisher = {American Physical Society},
  issn = {1550-7998, 1550-2368},
  doi = {10.1103/PhysRevD.71.014030},
  urldate = {2024-06-19},
  archiveprefix = {arXiv},
  langid = {english},
}

@article{Liu:2025bjm,
  title = {Dipionic Transitions of ${{Y}}(4500)$ to ${{J}}/\psi$},
  author = {Liu, Shidong and Wu, Qi and Li, Gang},
  year = 2025,
  month = oct,
  journal = {Phys. Rev. D},
  volume = {112},
  number = {7},
  eprint = {2504.14792},
  primaryclass = {hep-ph},
  pages = {074036},
  doi = {10.1103/fz1m-3mdg},
  urldate = {2025-05-06},
  archiveprefix = {arXiv},
}

@article{E852:2004rfa,
    author = "Lu, M. and others",
    collaboration = "E852",
    title = "{Exotic meson decay to $\omega\pi^0\pi^-$}",
    eprint = "hep-ex/0405044",
    archivePrefix = "arXiv",
    reportNumber = "JLAB-PHY-04-20",
    doi = "10.1103/PhysRevLett.94.032002",
    journal = "Phys. Rev. Lett.",
    volume = "94",
    pages = "032002",
    year = "2005"
}

@article{Maon:2024ifr,
    author = "Maon, Runqiu",
    collaboration = "BESIII",
    title = "{Observation of isoscalar $1^{-+}$ spin-exotic state $\eta_1$(1855)}",
    doi = "10.1393/ncc/i2024-24182-0",
    journal = "Nuovo Cim. C",
    volume = "47",
    number = "4",
    pages = "182",
    year = "2024"
}

@article{Wang:2012mf,
  title = {Open charm effects in the explanation of the long-standing ``$\rho\pi$ puzzle''},
  author = {Wang, Qian and Li, Gang and Zhao, Qiang},
  year = 2012,
  journal = {Phys. Rev. D},
  volume = {85},
  number = {7},
  eprint = {1201.1681},
  primaryclass = {hep-ph},
  pages = {074015},
  publisher = {American Physical Society},
  doi = {10.1103/PhysRevD.85.074015},
  urldate = {2023-05-05},
  archiveprefix = {arXiv},
}

@article{Meyer:2015eta,
    author = "Meyer, C. A. and Swanson, E. S.",
    title = "{Hybrid Mesons}",
    eprint = "1502.07276",
    archivePrefix = "arXiv",
    primaryClass = "hep-ph",
    doi = "10.1016/j.ppnp.2015.03.001",
    journal = "Prog. Part. Nucl. Phys.",
    volume = "82",
    pages = "21--58",
    year = "2015"
}

@article{E852:1998mbq,
    author = "Adams, G. S. and others",
    collaboration = "E852",
    title = "{Observation of a new $J^{PC} = 1^{-+}$ exotic state in the reaction $\pi^- p \to \pi^+ \pi^- \pi^- p$ at 18-GeV/$c$}",
    doi = "10.1103/PhysRevLett.81.5760",
    journal = "Phys. Rev. Lett.",
    volume = "81",
    pages = "5760--5763",
    year = "1998"
}

@article{Deandrea:2003pv,
  title = {{$J/\ensuremath{\psi}$} couplings to charmed resonances and to $\ensuremath{\pi}$},
  author = {Deandrea, A. and Nardulli, G. and Polosa, A. D.},
  year = 2003,
  journal = {Phys. Rev. D},
  volume = {68},
  number = {3},
  eprint = {hep-ph/0302273},
  pages = {034002},
  publisher = {American Physical Society},
  doi = {10.1103/PhysRevD.68.034002},
  urldate = {2023-04-07},
  archiveprefix = {arXiv},
}

@article{Mehen:2015efa,
  title = {Hadronic loops versus factorization in effective field theory calculations of $(3872)\to \chi_{cJ} \pi^0$},
  author = {Mehen, Thomas},
  year = 2015,
  month = aug,
  journal = {Phys. Rev. D},
  volume = {92},
  number = {3},
  eprint = {1503.02719},
  primaryclass = {hep-ph},
  pages = {034019},
  issn = {1550-7998, 1550-2368},
  doi = {10.1103/PhysRevD.92.034019},
  urldate = {2023-10-19},
  archiveprefix = {arXiv},
  langid = {english},
}

@article{Colangelo:2003sa,
  title = {Nonfactorizable contributions in B decays to charmonium: The case of ${B}^{\ensuremath{-}}\ensuremath{\rightarrow}{K}^{\ensuremath{-}}{h}_{c}$},
  shorttitle = {Nonfactorizable contributions in B decays to charmonium},
  author = {Colangelo, P. and De Fazio, F. and Pham, T.N.},
  year = 2004,
  journal = {Phys. Rev. D},
  volume = {69},
  number = {5},
  eprint = {hep-ph/0310084},
  pages = {054023},
  publisher = {American Physical Society},
  doi = {10.1103/PhysRevD.69.054023},
  urldate = {2024-01-30},
  archiveprefix = {arXiv},
}

@article{Amsler:1995td,
  title = {Is ${\mathit{f}}_{0}$(1500) a scalar glueball?},
  author = {Amsler, Claude and Close, Frank E.},
  year = 1996,
  journal = {Phys. Rev. D},
  volume = {53},
  number = {1},
  eprint = {hep-ph/9507326},
  pages = {295-311},
  publisher = {American Physical Society},
  doi = {10.1103/PhysRevD.53.295},
  urldate = {2023-05-05},
  archiveprefix = {arXiv},
}

@article{Huang:2022tpq,
    author = "Huang, Yin and Zhu, Hong Qiang",
    title = "{Revealing the inner structure of the newly observed ${\eta}_1(1855)$ via photoproduction}",
    eprint = "2209.02879",
    archivePrefix = "arXiv",
    primaryClass = "hep-ph",
    doi = "10.1088/1361-6471/ace4e2",
    journal = "J. Phys. G",
    volume = "50",
    number = "9",
    pages = "095002",
    year = "2023"
}

@article{Casalbuoni:1996pg,
  title = {Phenomenology of heavy meson chiral lagrangians},
  author = {Casalbuoni, R. and Deandrea, A. and Di Bartolomeo, N. and Gatto, R. and Feruglio, F. and Nardulli, G.},
  year = 1997,
  journal = {Phys. Rept.},
  volume = {281},
  number = {3},
  eprint = {hep-ph/9605342},
  pages = {145-238},
  issn = {0370-1573},
  doi = {10.1016/S0370-1573(96)00027-0},
  urldate = {2023-03-22},
  archiveprefix = {arXiv},
  langid = {english},
}

@article{ParticleDataGroup:2024cfk,
    author = "Navas, S. and others",
    collaboration = "Particle Data Group",
    title = "{Review of particle physics}",
    doi = "10.1103/PhysRevD.110.030001",
    journal = "Phys. Rev. D",
    volume = "110",
    number = "3",
    pages = "030001",
    year = "2024"
}

@article{Klempt:2007cp,
  title = {Glueballs, Hybrids, Multiquarks. Experimental facts versus QCD inspired concepts},
  author = {Klempt, Eberhard and Zaitsev, Alexander},
  year = 2007,
  journal = {Phys. Rept.},
  volume = {454},
  eprint = {0708.4016},
  primaryclass = {hep-ph},
  pages = {1-202},
  doi = {10.1016/j.physrep.2007.07.006},
  archiveprefix = {arXiv},
}

@article{BESIII:2022iwi,
  title = {Partial Wave Analysis of $J/\psi\to\gamma\eta\eta^\prime$},
  author = {Ablikim, M. and others},
  year = 2022,
  month = oct,
  journal = {Phys. Rev. D},
  volume = {106},
  number = {7},
  eprint = {2202.00623},
  primaryclass = {hep-ex},
  pages = {072012},
  doi = {10.1103/PhysRevD.106.072012},
  archiveprefix = {arXiv},
  collaboration = {BESIII},
}

@article{Chen:2022qpd,
  title = {{{QCD}} Axial Anomaly Enhances the $\eta\eta^\prime$ Decay of the Hybrid Candidate $\eta_1(1855)$},
  author = {Chen, Hua-Xing and Su, Niu and Zhu, Shi-Lin},
  year = 2022,
  journal = {Chin. Phys. Lett.},
  volume = {39},
  number = {5},
  eprint = {2202.04918},
  primaryclass = {hep-ph},
  pages = {051201},
  issn = {0256-307X, 1741-3540},
  doi = {10.1088/0256-307X/39/5/051201},
  urldate = {2025-10-11},
  archiveprefix = {arXiv},
}

@article{COMPASS:2021ogp,
    author = "Alexeev, M. G. and others",
    collaboration = "COMPASS",
    title = "{Exotic meson $\pi_1(1600)$ with $J^{PC} = 1^{-+}$ and its decay into $\rho(770)\pi$}",
    eprint = "2108.01744",
    archivePrefix = "arXiv",
    primaryClass = "hep-ex",
    reportNumber = "CERN-EP-2021{\textendash}162",
    doi = "10.1103/PhysRevD.105.012005",
    journal = "Phys. Rev. D",
    volume = "105",
    number = "1",
    pages = "012005",
    year = "2022"
}

@article{Esmer:2025xss,
    author = {Esmer, G. Daylan and Azizi, K. and Sundu, H. and T{\"u}rkmen, S.},
    title = "{Decays of the light hybrid meson $1^{-+}$}",
    eprint = "2501.11331",
    archivePrefix = "arXiv",
    primaryClass = "hep-ph",
    doi = "10.1103/PhysRevD.111.034041",
    journal = "Phys. Rev. D",
    volume = "111",
    number = "3",
    pages = "034041",
    year = "2025"
}

@article{Dong:2022cuw,
  title = {Interpretation of the {$\eta_1(1855)$} as a ${K}\bar {K}_1(1400)+$ c.c. Molecule},
  author = {Dong, Xiang-Kun and Lin, Yong-Hui and Zou, Bing-Song},
  year = 2022,
  month = may,
  journal = {Sci. China Phys. Mech. Astron.},
  volume = {65},
  number = {6},
  eprint = {2202.00863},
  primaryclass = {hep-ph},
  pages = {261011},
  doi = {10.1007/s11433-022-1887-5},
  urldate = {2025-10-11},
  archiveprefix = {arXiv},
}

@article{DM2:1988bfq,
  title = "{$J/\psi \to \mathrm{vector}+\mathrm{pseudoscalar}$ decays and the quark content of $\ensuremath{\eta}$ and ${\ensuremath{\eta}}^{\prime}$}",
  author = {Jousset, J. and others},
  year = 1990,
  journal = {Phys. Rev. D},
  volume = {41},
  number = {5},
  pages = {1389},
  publisher = {American Physical Society},
  doi = {10.1103/PhysRevD.41.1389},
  urldate = {2023-05-05},
  collaboration = {DM2},
}

@article{Yu:2022lwl,
    author = "Yu, Yao and Xiong, Zhuang and Zhang, Han and Ke, Bai-Cian and Teng, Yi and Liu, Qing-Shan and Zhang, Jia-Wei",
    title = "{Investigating the ${\eta}_1'(1855)$ exotic state in the $J/\psi \to\eta_1'(1855)\eta^{(\prime)}$ decays}",
    eprint = "2208.05442",
    archivePrefix = "arXiv",
    primaryClass = "hep-ph",
    doi = "10.1016/j.physletb.2023.137965",
    journal = "Phys. Lett. B",
    volume = "842",
    pages = "137965",
    year = "2023"
}

@article{JPAC:2018zyd,
    author = "Rodas, A. and others",
    collaboration = "JPAC",
    title = "{Determination of the pole position of the lightest hybrid meson candidate}",
    eprint = "1810.04171",
    archivePrefix = "arXiv",
    primaryClass = "hep-ph",
    reportNumber = "JLAB-THY-18-2839",
    doi = "10.1103/PhysRevLett.122.042002",
    journal = "Phys. Rev. Lett.",
    volume = "122",
    number = "4",
    pages = "042002",
    year = "2019"
}

@article{Alde:1991qz,
    author = "Alde, D. and others",
    title = "{Further study of the $X (1910)$ meson}",
    reportNumber = "IFVE-91-40",
    journal = "Sov. J. Nucl. Phys.",
    volume = "54",
    pages = "455--458",
    year = "1991"
}

@article{E852:2004gpn,
    author = "Kuhn, Joachim and others",
    collaboration = "E852",
    title = "{Exotic meson production in the $f_1(1285) \pi^-$ system observed in the reaction $\pi^- p\to \eta\pi^+\pi^-\pi^-$ at 18 GeV/c}",
    eprint = "hep-ex/0401004",
    archivePrefix = "arXiv",
    doi = "10.1016/j.physletb.2004.05.032",
    journal = "Phys. Lett. B",
    volume = "595",
    pages = "109--117",
    year = "2004"
}

@article{Shastry:2023ths,
  title = {Radiative Production and Decays of the Exotic {$\eta_1^\prime(1855)$} and Its Siblings},
  author = {Shastry, Vanamali and Giacosa, Francesco},
  year = 2023,
  month = may,
  journal = {Nucl. Phys. A},
  volume = {1037},
  eprint = {2302.07687},
  primaryclass = {hep-ph},
  pages = {122683},
  issn = {03759474},
  doi = {10.1016/j.nuclphysa.2023.122683},
  urldate = {2025-10-11},
  archiveprefix = {arXiv},
}

@article{Dudek:2013yja,
  title = {Toward the Excited Isoscalar Meson Spectrum from Lattice {{QCD}}},
  author = {Dudek, Jozef J. and Edwards, Robert G. and Guo, Peng and Thomas, Christopher E.},
  year = 2013,
  month = nov,
  journal = {Phys. Rev. D},
  volume = {88},
  number = {9},
  eprint = {1309.2608},
  primaryclass = {hep-lat},
  pages = {094505},
  issn = {1550-7998, 1550-2368},
  doi = {10.1103/PhysRevD.88.094505},
  urldate = {2025-06-30},
  archiveprefix = {arXiv},
  collaboration = {Hadron Spectrum},
}

@article{Zhang:2025xee,
    author = "Zhang, Fu-Yuan and Huang, Qi and Wang, Li-Ming",
    title = "{Spectral analysis and decay mechanisms of $1^{-+}$ hybrid states in light meson sector}",
    eprint = "2503.01443",
    archivePrefix = "arXiv",
    primaryClass = "hep-ph",
    doi = "10.1103/1wkx-3s2l",
    journal = "Phys. Rev. D",
    volume = "113",
    number = "1",
    pages = "014002",
    year = "2026"
}

@article{Yang:2022rck,
  title = {Analysis of the {$\eta_1(1855)$} as a ${K}\bar {K}_1(1400)$ Molecular State},
  author = {Yang, Feng and Zhu, Hong Qiang and Huang, Yin},
  year = 2023,
  month = feb,
  journal = {Nucl. Phys. A},
  volume = {1030},
  eprint = {2203.06934},
  primaryclass = {hep-ph},
  pages = {122571},
  issn = {03759474},
  doi = {10.1016/j.nuclphysa.2022.122571},
  urldate = {2025-10-11},
  archiveprefix = {arXiv},
}

@misc{BESIII:2025vea,
    author = "Ablikim, Medina and others",
    collaboration = "BESIII",
    title = "{Search for a $1^{-+}$ molecular state via $e^{+}e^{-} \to \gamma D^{+}_{s} D_{s1}^{-}(2536) +c.c.$}",
    eprint = "2503.11015",
    archivePrefix = "arXiv",
    primaryClass = "hep-ex",
    month = "3",
    year = "2025"
}

@article{Amelin:2005ry,
    author = "Amelin, D. V. and others",
    editor = "Abov, Yu. G.",
    title = "{Investigation of hybrid states in the VES experiment at the Institute for High Energy Physics (Protvino)}",
    doi = "10.1134/1.1891185",
    journal = "Phys. Atom. Nucl.",
    volume = "68",
    pages = "359--371",
    year = "2005"
}

@article{Bai:2022cfz,
  title = {$\mathrm{\ensuremath{\Upsilon}}(10753)\ensuremath{\rightarrow}\mathrm{\ensuremath{\Upsilon}}(\mathrm{n}\mathrm{S}){\ensuremath{\pi}}^{+}{\ensuremath{\pi}}^{\ensuremath{-}}$ decays induced by hadronic loop mechanism},
  author = {Bai, Zi-Yue and Li, Yu-Shuai and Huang, Qi and Liu, Xiang and Matsuki, Takayuki},
  year = 2022,
  month = apr,
  journal = {Phys. Rev. D},
  volume = {105},
  number = {7},
  eprint = {2201.12715},
  primaryclass = {hep-ph},
  pages = {074007},
  publisher = {American Physical Society},
  doi = {10.1103/PhysRevD.105.074007},
  urldate = {2024-01-30},
  archiveprefix = {arXiv},
}

@article{Colangelo:2002mj,
  title = "{$B^-\to K^-\chi_{c0}$ decay from charmed meson rescattering}",
  author = {Colangelo, P. and De Fazio, F. and Pham, T. N.},
  year = 2002,
  journal = {Phys. Lett. B},
  volume = {542},
  number = {1},
  eprint = {hep-ph/0207061},
  pages = {71-79},
  issn = {0370-2693},
  doi = {10.1016/S0370-2693(02)02306-7},
  urldate = {2023-08-03},
  archiveprefix = {arXiv},
  langid = {english},
}

@misc{BESIII:2025bez,
    author = "Ablikim, Medina and others",
    collaboration = "BESIII",
    title = "{Search for $1^{-+}$ charmonium-like hybrid via $e^{+}e^{-}\rightarrow \gamma \eta^{(\prime)} \eta_{c}$ at center-of-mass energies between 4.258 and 4.681 GeV}",
    eprint = "2504.13539",
    archivePrefix = "arXiv",
    primaryClass = "hep-ex",
    month = "4",
    year = "2025"
}

@article{BESIII:2022riz,
  title = "{Observation of an  Isoscalar Resonance  with  Exotic $J^{PC}=1^{-+}$  Quantum Numbers  in $J/\psi\to\gamma\eta\eta^\prime$}",
  author = "Ablikim, M. and others",
  collaboration = "BESIII",
  eprint = "2202.00621",
  archivePrefix = "arXiv",
  primaryClass = "hep-ex",
  doi = "10.1103/PhysRevLett.129.192002",
  journal = "Phys. Rev. Lett.",
  volume = "129",
  number = "19",
  pages = "192002",
  year = "2022",
  note = "[Erratum: Phys.Rev.Lett. 130, 159901 (2023)]"
}

@article{Wang:2022sib,
  title = {Production of the {$\eta_1(1855)$} through Kaon Induced Reactions under the Assumptions That It Is a Molecular or a Hybrid State},
  author = {Wang, Xiao-Yun and Zeng, Fan-Cong and Liu, Xiang},
  year = 2022,
  month = aug,
  journal = {Phys. Rev. D},
  volume = {106},
  number = {3},
  eprint = {2205.09283},
  primaryclass = {hep-ph},
  pages = {036005},
  issn = {2470-0010, 2470-0029},
  doi = {10.1103/PhysRevD.106.036005},
  urldate = {2025-10-11},
  archiveprefix = {arXiv},
}

@article{Jia:2025xil,
    author = "Jia, Zhao-Sai and Li, Gang and Zhang, Zhen-Hua",
    title = "{Constrain the $\chi_{cJ}\to D^{(\ast)}\bar{D}^{(\ast)}$ effective couplings via the $X(3872)\to \chi_{cJ} \pi^0$ decays}",
    eprint = "2507.16618",
    archivePrefix = "arXiv",
    primaryClass = "hep-ph",
    doi = "10.1103/51z1-qls5",
    journal = "Phys. Rev. D",
    volume = "112",
    number = "7",
    pages = "074035",
    year = "2025"
}

@article{Chen:2015bma,
  title = "{Search for missing $\ensuremath{\psi}(4S)$ in the ${e}^{+}{e}^{\ensuremath{-}}\ensuremath{\rightarrow}{\ensuremath{\pi}}^{+}{\ensuremath{\pi}}^{\ensuremath{-}}\ensuremath{\psi}(2S)$ process}",
  author = {Chen, Dian-Yong and Liu, Xiang and Matsuki, Takayuki},
  year = 2016,
  month = feb,
  journal = {Phys. Rev. D},
  volume = {93},
  number = {3},
  eprint = {1509.00736},
  primaryclass = {hep-ph},
  pages = {034028},
  publisher = {American Physical Society},
  doi = {10.1103/PhysRevD.93.034028},
  urldate = {2023-11-21},
  archiveprefix = {arXiv},
}

@article{Guo:2010ak,
  title = {Effect of charmed meson loops on charmonium transitions},
  author = {Guo, Feng-Kun and Hanhart, C. and Li, Gang and Meißner, Ulf-G. and Zhao, Qiang},
  year = 2011,
  journal = {Phys. Rev. D},
  volume = {83},
  number = {3},
  eprint = {1008.3632},
  primaryclass = {hep-ph},
  pages = {034013},
  publisher = {American Physical Society},
  doi = {10.1103/PhysRevD.83.034013},
  urldate = {2023-02-13},
  archiveprefix = {arXiv},
}

@article{Chen:2022isv,
  title = "{$1^{-+}$ Hybrid Meson in $J/\psi$ Radiative Decays from Lattice {{QCD}}}",
  author = {Chen, Feiyu and Jiang, Xiangyu and Chen, Ying and Gong, Ming and Liu, Zhaofeng and Shi, Chunjiang and Sun, Wei},
  year = 2023,
  month = mar,
  journal = {Phys. Rev. D},
  volume = {107},
  number = {5},
  eprint = {2207.04694},
  primaryclass = {hep-lat},
  pages = {054511},
  publisher = {American Physical Society},
  doi = {10.1103/PhysRevD.107.054511},
  urldate = {2025-06-30},
  archiveprefix = {arXiv},
}

@article{Baker:2003jh,
    author = "Baker, C. A. and others",
    title = "{Confirmation of $a_0(1450)$ and $\pi_1(1600)$ in $\bar p p\to \omega \pi^+ \pi^- \pi^0$ at rest}",
    doi = "10.1016/S0370-2693(03)00643-9",
    journal = "Phys. Lett. B",
    volume = "563",
    pages = "140--149",
    year = "2003"
}

@article{COMPASS:2014vkj,
    author = "Adolph, C. and others",
    collaboration = "COMPASS",
    title = "{Odd and even partial waves of $\eta\pi^-$ and $\eta'\pi^-$ in $\pi^-p\to\eta^{(\prime)}\pi^-p$ at $191\,\textrm{GeV}/c$}",
    eprint = "1408.4286",
    archivePrefix = "arXiv",
    primaryClass = "hep-ex",
    reportNumber = "CERN-PH-EP-2014-204",
    doi = "10.1016/j.physletb.2014.11.058",
    journal = "Phys. Lett. B",
    volume = "740",
    pages = "303--311",
    year = "2015",
    note = "[Erratum: Phys.Lett.B 811, 135913 (2020)]"
}

@article{Wu:2025crk,
    author = "Wu, Qi and Sun, Zhong-Quan and Chen, Dian-Yong and Liu, Shi-Dong and Li, Gang",
    title = "{Dipion transitions from $X(3872)$ to $\chi_{cJ}\ (J=0,1,2)$}",
    eprint = "2512.21161",
    archivePrefix = "arXiv",
    primaryClass = "hep-ph",
    doi = "10.1103/19lr-96hx",
    journal = "Phys. Rev. D",
    volume = "113",
    number = "1",
    pages = "014015",
    year = "2026"
}

@article{Dudek:2010wm,
    author = "Dudek, Jozef J. and Edwards, Robert G. and Peardon, Michael J. and Richards, David G. and Thomas, Christopher E.",
    title = "{Toward the excited meson spectrum of dynamical QCD}",
    eprint = "1004.4930",
    archivePrefix = "arXiv",
    primaryClass = "hep-ph",
    reportNumber = "JLAB-THY-10-1171",
    doi = "10.1103/PhysRevD.82.034508",
    journal = "Phys. Rev. D",
    volume = "82",
    pages = "034508",
    year = "2010"
}

@article{VES:1993scg,
    author = "Beladidze, G. M. and others",
    collaboration = "VES",
    title = "{Study of $\pi^- N \to \eta \pi^- N$ and $\pi^- N \to \eta^\prime \pi^- N$ reactions at 37 GeV/$c$}",
    reportNumber = "IFVE-93-62",
    doi = "10.1016/0370-2693(93)91224-B",
    journal = "Phys. Lett. B",
    volume = "313",
    pages = "276--282",
    year = "1993"
}

@article{Shastry:2022mhk,
  title = {The Phenomenology of the Exotic Hybrid Nonet with {$\pi_1(1600)$} and {$\eta_1(1855)$}},
  author = {Shastry, Vanamali and Fischer, Christian S. and Giacosa, Francesco},
  year = 2022,
  month = oct,
  journal = {Phys. Lett. B},
  volume = {834},
  eprint = {2203.04327},
  primaryclass = {hep-ph},
  pages = {137478},
  issn = {03702693},
  doi = {10.1016/j.physletb.2022.137478},
  urldate = {2025-10-11},
  archiveprefix = {arXiv},
}

@article{Tan:2025ahx,
    author = "Tan, Yue and Wu, Yu-Heng and Huang, Qi and Chen, Xiaoyun and Hu, Xiaohuang and Yang, Youchang and Ping, Jialun",
    title = "{Absence of ${\eta}_1(1855)$ and its exotic partner candidate in a chiral quark model}",
    eprint = "2504.10924",
    archivePrefix = "arXiv",
    primaryClass = "hep-ph",
    doi = "10.1103/nw3m-tzrb",
    journal = "Phys. Rev. D",
    volume = "112",
    number = "9",
    pages = "094004",
    year = "2025"
}
\end{document}